%% file: Journal_Regular_CF_Massive_SwitchOnOff.tex
\documentclass[journal]{IEEEtran}

\usepackage[latin1]{inputenc}
\usepackage{times}
\usepackage{amsmath}
\usepackage{amssymb}
\usepackage{amsthm}
\usepackage{mathrsfs}
\usepackage{subcaption}
\usepackage{graphicx}
\usepackage{enumerate}
\usepackage[compress]{cite}
\usepackage{float}
\usepackage[acronym]{glossaries}
\usepackage{enumitem}
\usepackage{algorithmic}
\usepackage{algorithm}
\usepackage{multirow}
\usepackage{etoolbox}
\usepackage[latin1]{inputenc}
\usepackage{times}
\usepackage{amsmath}
\usepackage{amssymb}
\usepackage{amsthm}
\usepackage{mathrsfs}
\usepackage{subcaption}
\usepackage{graphicx}
\usepackage{enumerate}
\usepackage[compress]{cite}
\usepackage{float}
\usepackage[acronym]{glossaries}
\usepackage{enumitem}
\usepackage{algorithmic}
\usepackage{algorithm}
\usepackage{multirow}
\usepackage{etoolbox}
\usepackage{subcaption}
\usepackage{lipsum}
\usepackage{mathtools}
\usepackage{cuted}
\usepackage{subcaption}

\def\BibTeX{{\rm B\kern-.05em{\sc i\kern-.025em b}\kern-.08em
    T\kern-.1667em\lower.7ex\hbox{E}\kern-.125emX}}

\input{acronyms}

\newcommand{\bs}{\boldsymbol}
\DeclareMathOperator*{\argmax}{arg\,max}
\DeclareMathOperator*{\argmin}{arg\,min}

\begin{document}

%\title{Cell-free principles on cellular topologies: spectral and energy efficiency aspects
%\thanks{We acknowledge support from grants IRENE-STARMAN (PID2020-115323RB-C32 funded by MCIN/AEI/10.13039/501100011033, Spain), and GERMINAL (TED2021-131624B-I00, funded by MCIN/AEI/10.13039/ 501100011033, Spain, and by European Union ''NextGenerationEU''/PRTR).}
%}

\title{Energy efficient cell-free massive MIMO on 5G deployments: sleep modes and user stream management
\thanks{We acknowledge support from grants IRENE-STARMAN (PID2020-115323RB-C32 funded by MCIN/AEI/10.13039/501100011033, Spain), GERMINAL (TED2021-131624B-I00, funded by MCIN/AEI/10.13039/ 501100011033, Spain, and by European Union ''NextGenerationEU''/PRTR), and iTENTE (CIDEXG/2022/17, funded by CIDEGENT PlaGenT, Generalitat Valenciana, Spain).}
}

%\author{
%    \IEEEauthorblockN{F. Riera-Palou\IEEEauthorrefmark{1}, G. Femenias\IEEEauthorrefmark{1}, D. Lopez-Perez\IEEEauthorrefmark{2}, N. Piovesan\IEEEauthorrefmark{2} and A. de Domenico\IEEEauthorrefmark{2}}\\
%    \IEEEauthorblockA{\IEEEauthorrefmark{1}Mobile Communications Group - Universitat de les Illes Balears - 07122 Mallorca, Spain}\\
%    \IEEEauthorblockA{\IEEEauthorrefmark{2}Huawei Technologies - 92100 Bolougne-Billancourt, France}
%    \\Email: \{felip.riera,guillem.femenias\}@uib.es, \{david.lopez.perez,nicola.piovesan,antonio.de.domenico\}@huawei.com
%}

\author{Felip~Riera-Palou,~\IEEEmembership{Senior Member,~IEEE,}
        Guillem~Femenias,~\IEEEmembership{Senior Member,~IEEE,}
        David~Lopez-Perez,~\IEEEmembership{Senior Member,~IEEE,}
        Nicola~Piovesan,~\IEEEmembership{Senior Member,~IEEE,}
        Antonio~de~Domenico,~\IEEEmembership{Senior Member,~IEEE}
\thanks{F. Riera-Palou and G. Femenias are with the Mobile Communications Group, University of the Balearic Islands, Palma 07122, Illes Balears, Spain (e-mail: \{felip.riera,guillem.femenias\}@uib.es). D Lopez-Perez is with
Universitat Polit\`ecnica de Val\`encia - 46022 Val\`encia, Spain (e-mail: d.lopez@iteam.upv.es).
 N. Piovesan and A. De Domenico are with Huawei Technologies, 92100 Bolougne-Billancourt, France \{nicola.piovesan,antonio.de.domenico\}@huawei.com.}
%\thanks{Manuscript received September 15, 2016.}
}

\maketitle

\title{Energy efficient cell-free massive MIMO on 5G deployments: sleep modes strategies and user stream management
\thanks{We acknowledge support from grants IRENE-STARMAN (PID2020-115323RB-C32 funded by MCIN/AEI/10.13039/501100011033, Spain), GERMINAL (TED2021-131624B-I00, funded by MCIN/AEI/10.13039/ 501100011033, Spain, and by European Union ''NextGenerationEU''/PRTR), and iTENTE (CIDEXG/2022/17, funded by CIDEGENT PlaGenT, Generalitat Valenciana, Spain).}
}

\author{
    \IEEEauthorblockN{F. Riera-Palou\IEEEauthorrefmark{1}, G. Femenias\IEEEauthorrefmark{1}, D. Lopez-Perez\IEEEauthorrefmark{2}, N. Piovesan\IEEEauthorrefmark{3} and A. De Domenico\IEEEauthorrefmark{3}\thanks{We acknowledge support from grants IRENE-STARMAN (PID2020-115323RB-C32 funded by MCIN/AEI/10.13039/501100011033, Spain), GERMINAL (TED2021-131624B-I00, funded by MCIN/AEI/10.13039/ 501100011033, Spain, and by European Union ''NextGenerationEU''/PRTR), and iTENTE (CIDEXG/2022/17, funded by CIDEGENT PlaGenT, Generalitat Valenciana, Spain).}}
    \IEEEauthorblockA{\IEEEauthorrefmark{1}Mobile Communications Group - Universitat de les Illes Balears - 07122 Mallorca, Spain}
    \IEEEauthorblockA{\IEEEauthorrefmark{2}Universitat Polit\`ecnica de Val\`encia - 46022 Val\`encia, Spain}
    \IEEEauthorblockA{\IEEEauthorrefmark{3}Huawei Technologies - 92100 Bolougne-Billancourt, France}
    \\Email: \{felip.riera,guillem.femenias\}@uib.es, d.lopez@iteam.upv.es, \{nicola.piovesan,antonio.de.domenico\}@huawei.com
}

\maketitle

\begin{abstract}
This paper proposes the utilization of cell-free massive MIMO (CF-M-MIMO) processing on top of the regular micro/macrocellular deployments typically found in current 5G networks.
Towards this end, it contemplates the connection of all base stations to a central processing unit (CPU) through fronthaul links, thus enabling the joint processing of all serviced user equipment (UE),
yet avoiding the expensive deployment and maintenance of dozens of randomly scattered access points (APs). Moreover, it allows the implementation of centralized strategies to exploit the sleep mode capabilities of current baseband/RF hardware to (de)activate selected Base Stations (BSs) in order to maximize the network energy efficiency and to react to changes in UE behaviour and/or operator requirements. In line with current cellular network deployments, it considers the use of multiple antennas at the UE side that unavoidably introduces the need to effectively manage the number of streams to be directed to each UE in order to balance multiplexing gains and increased pilot contamination.
\end{abstract}

\begin{IEEEkeywords}
 Cell-free, Massive MIMO, Multicell, Energy efficiency, UE-centric, Mutiple-antenna UE.
\end{IEEEkeywords}

\input{acronyms}
\section{Introduction}

Research on the next-generation of mobile networks ---so-called 6G--- is well underway in academy, industry and standardization bodies \cite{jiang2021,matthaiou2021,tataria2021}.
Despite the many unknowns yet,
a few technological trends seem bound to have a central role to support a host of new applications such as augmented reality or autonomous driving~\cite{uusitalo2021}.
In particular,
the exploitation of new frequency bands (e.g., mmWave and THz) and networking elements (e.g., reconfigurable intelligent surfaces),
the integration of a non-terrestrial segment,
as well as the introduction of new enhancements,
which expand the capabilities of current features (e.g., advanced \gls{M-MIMO}, sleep modes),
are invariably cited as key 6G enablers.
Among them,
enhancements to \gls{M-MIMO} \cite{marzetta2016},
whose first incarnations are already embedded in current 5G networks,
are being intensively investigated to expand 6G capabilities in terms of coverage, capacity and reliability.
One promising new flavour of \gls{M-MIMO} ---namely \gls{CF-M-MIMO}--- combines the utilization of a massive number of antennas with the principles underpinning small cell networks \cite{lopez2022}.
In \gls{CF-M-MIMO}, 
the antennas are distributed throughout the coverage area using a multitude of \glspl{TRP}, 
each one of them equipped with a modest number of antennas,
and connected to a single \gls{CPU} to benefit from a centralised joint processing \cite{ngo2017} 
(see \cite{demir2021} for an state-of-the-art survey).

%Initially proposed in \cite{ngo2017} to provide equal QoS (i.e., equal rates) to all \glspl{UE}, \gls{CF-M-MIMO} assumes the existence of a single \gls{CPU} to which an abundance of \glspl{TRP}, irregularly distributed over the area to be covered, are connected via fronthaul links. %The first \gls{CF-M-MIMO} proposals relied on  distributed precoding typically implemented using local \gls{CB} processing at each \gls{TRP}. Subsequently, \cite{nayebi2017} showed the great performance benefits that could be attained using CPU-based centralized precoding although this came at the cost of having to centralize the precoder design at the CPU, and thus requiring the exchange of fast-fading instantaneous information (e.g., channel gains, precoding vectors) over the CPU-TRP fronthaul links \cite{femenias2019}.
%Over the last few years, a plethora of precoding and power/pilot allocation techniques have been proposed with a variety of performance trade-offs (see \cite{demir2021} for an state-of-the-art survey).

A common challenge to all these new developments in cellular networks ---and particularly to \gls{CF-M-MIMO}--- is that of network \gls{EE} \cite{lopez2022}.
The unprecedented number of nodes and fronthaul connections required to realize a \gls{CF-M-MIMO} network may significantly increase the network energy consumption,
and in turn,
the operators' CO$_2$ emissions and electricity bills.
To witness the success of \gls{CF-M-MIMO},
the research community will need to overcome this \gls{EE} challenge.

\gls{CF-M-MIMO} networks have recently been examined from an \gls{EE} perspective.
From an optimization standpoint,
the work in \cite{mai2022} (and references therein) stands out,
proposing different techniques to optimize the transmit power allocation to maximize \gls{EE}.
A rather different approach is introduced in \cite{femenias2020,garcia2020},
in which the network \gls{EE} is maximized by taking advantage of sleep modes and selectively (de)activating \glspl{TRP}.
%It has been recently shown that large gains in \gls{EE} can be achieved in practical 5G networks by switching on/off different parts of the \glspl{TRP} transceiver chains \cite{lopez2022}.
Despite all the promised \gls{EE} benefits,
the vanilla \gls{CF-M-MIMO} proposed in the literature may still be hindered by practical constraints,
mostly arising from the enormous cost of deploying and maintaining a large number of \glspl{TRP} and their corresponding connections to the \gls{CPU} through fronthaul links \cite{kanno2022}.
Coordinating sleep modes in such small cell scenario setup is also challenging.
Consequently,
intermediate solutions combining a macrocellular 5G network with small-scale \gls{CF-M-MIMO} deployments have been recently investigated  from an \gls{EE} viewpoint \cite{kim2022}.

In this paper, 
we take an even more cautious and practical approach,
considering the application of \gls{CF-M-MIMO} principles to a state-of-the-art 5G macrocellular network taking into account practical aspects of both, 
the macrocellular infrastructure and off-the-shelf \glspl{UE}.
In particular, we investigate for the first time the \gls{EE} of a practical 5G macrocellular \gls{CF-M-MIMO} network, 
where all the macro \glspl{TRP}, 
each equipped with a \gls{M-MIMO} array, 
are connected to a central CPU through fronthaul links (no additional randomly scattered \glspl{TRP} are required/considered) to potentially enable the joint processing of all \glspl{UE} in the coverage area.
In this setup,
the \glspl{TRP} act as \glspl{AP} in cell-free terminology.
%Note that the key difference with respect to legacy cooperation mechanisms such as \gls{CoMP} is the superior degree of joint processing flexibility enabled by the merging of all \glspl{UE} and \glspl{TRP} information at a central node (\gls{CoMP} is often limited to coordinate three adjacent sectors).
Contrary to most of the \gls{CF-M-MIMO} literature, 
we also consider multi-antenna \glspl{UE}, 
instead of single-antenna ones,
as the vast majority of off-the-shelf \glspl{UE} in use today are equipped with multiple antennas. 
Multiple-antennas at the \gls{UE} side in the context of downlink \gls{CF-M-MIMO} to allow single-\gls{UE} stream multiplexing were considered in \cite{mai2020}. 
However, this study was conducted in the absence of pilot contamination effects,
which will be shown later on in this paper to have a large impact on \gls{SE}. 
Provided this setup, 
in this paper, we investigate adaptive \gls{TRP} and multi-stream (de)activation to achieve a much more flexible approach to \gls{EE} and \gls{SE} management,
specifically tailored to currently deployed 5G infrastructure. 
In more detail, 
the main contributions of this paper are thus as follows:
\begin{enumerate}
    \item  
    The proposal and assessment of the application of \gls{CF-M-MIMO} centralized baseband processing to a regular \gls{M-MIMO} cellular topology,
    while taking into account scalability issues, general propagation conditions (i.e., indoor and outdoor \glspl{UE}), the presence of multiple-antenna at the \gls{UE} side and the availability of a state-of-the-art power consumption model. 
    In other words, we assess what the \gls{CF-M-MIMO} processing can bring along, 
    when applied under conditions typically encountered nowadays on currently deployed cellular infrastructure.
    \item 
    A heuristic technique to exploit the multiple-antenna processing capability of \glspl{UE},
    while relying on statistical channel knowledge, 
    is introduced. 
    As it will be shown, the naive and indiscriminate activation of multiple streams causes an abrupt increase in pilot contamination, 
    as each active \gls{UE} antenna requires a separate pilot to estimate the corresponding channel. 
    Furthermore, in the downlink, 
    per-stream power split can potentially lead to lower \glspl{SINR} at the receiving end. 
    Both effects, 
    if not handled properly, 
    severely hinder the performance of the weakest \glspl{UE} in the network. 
    The proposed technique aims at maintaining the rates of the worst \glspl{UE} in the network, 
    while allowing the strongest \glspl{UE} to attain higher rates by using multi-stream transmissions.
    \item 
    Relying on the sleep mode philosophy of \cite{lopez2022,femenias2020,garcia2020}, 
    and exploiting a novel consumption model for 5G \gls{M-MIMO} infrastructure,
    a new and flexible algorithm is introduced to improve the \gls{EE} of the network by selectively switching-off \gls{M-MIMO} \glspl{TRP}, 
    while subject to constraints on the \glspl{UE}' \gls{SE}. 
    These constraints can be formulated relying on the network sumrate, average \glspl{UE}' rate or 5\%-tile \glspl{UE}' rate.
    \item 
    An exhaustive numerical evaluation of the proposed techniques is presented in a framework largely compliant with industry standards \cite{itu2017}, 
    incorporating most traits characterizing 5G networks. 
    The effects of indoor propagation, the presence of multi-antenna \glspl{UE} or the impact of the \gls{ISD} are all assessed and discussed.
\end{enumerate}
Anticipating the results to be shown, 
it is envisaged that, 
due to the use of joint transmissions throughout the network, 
coverage is significantly enhanced, 
thus allowing a more aggressive shutdown of \gls{M-MIMO} \glspl{TRP}, 
unleashing \gls{EE} gains never exploited before.

\textbf{\emph{Notational remark}}:
Vectors and matrices are denoted by lower- and upper-case bold characters, respectively. 
The matrix operator, $\text{vec}(\bs A)$, stacks the columns of matrix $\bs A$ into a column vector. 
The Kronecker product of two matrices is denoted by operator $\otimes$, 
whereas $\|\bs A\|_F$ serves to represent the Frobenius norm of matrix $\bs A$. 
The identity matrix of dimension $L$ is denoted by $\bs I_L$, 
and an $M\times N$ matrix of zeros is represented by $\bs 0_{M\times N}$. 
The operator, $\mathcal D(\bs x)$, results in an squared diagonal matrix having $\bs x$ at its main diagonal. 
Similarly, when applied to a set of matrices, 
$\mathcal D(\bs X_1, \hdots, \bs X_n)$ results in a block-diagonal matrix with matrices $\bs X_1, \hdots, \bs X_n$ at its main diagonal. 
The expectation of a random scalar/vector/matrix variable is denoted by operator $E\{\cdot \}$.

\section{System model}
\label{sec:model}

\begin{figure}[!t]
      \begin{center}
       \includegraphics[width=.7\linewidth]{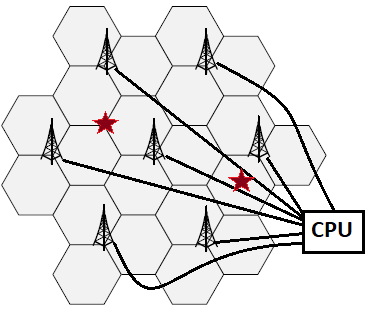}
      \end{center}
      \caption{Dense Urban scenario according to \cite{itu2017} with \gls{BS} sites coordinated through a \gls{CPU}.}% Circled \gls{UE} depicts a \gls{UE} that has been offloaded from the terrestrial to the satellite segment.}
      \label{topology}
      \vspace{-0.5cm}
\end{figure}

This paper considers a wireless network such as the one depicted in Fig.~\ref{topology},
whereby $L$ \gls{BS} sites are deployed in a regular fashion in accordance to the \gls{eMBB} dense urban scenario introduced in \cite{itu2017}.
Every \gls{BS} site is assumed to be composed of $S=3$ \glspl{TRP},
each of them providing a 120 degree sectorial coverage towards a certain direction by means of a \gls{M-MIMO} antenna array with $N_\text{TRP}$ antenna elements,
thus resulting in a total of $LSN_\text{TRP}$ antennas servicing the overall target area.
For convenience,
we denote by $M_T=LS$ the total number of \glspl{TRP} in the network.
Building on the \gls{C-RAN} concept \cite{checko2014} ---or in the more recent cell-free concept \cite{demir2021}---,
we consider that all \glspl{TRP} are connected to a \gls{CPU} (or group of \glspl{CPU}) by means of ideal fronthaul links.
Note that the well-planned placement of this macrocellular \glspl{TRP} throughout the coverage area is in sharp contrast to the vast majority of the cell-free literature
that tends to consider a random deployment of \glspl{TRP} according to the small cell philosophy.
The rationale behind the usage of the regular macrocellular deployment considered here has to be sought on the fact that,
at least for the first cell-free incarnations,
operators are more likely to heavily rely on the reuse of the existing well-planned site infrastructure to alleviate costs.

In this work, owing to its superior performance, 
centralized processing will be embraced.
However, for scalability purposes, each \gls{TRP} will only serve a maximum of $K_\text{TRP}$ \glspl{UE}.
This restriction ensures that, 
regardless of the total number of \glspl{UE} in the network,
the \glspl{TRP} hardware and fronthaul requirements remain bounded \cite{bjornson2020}.

This network exploits a bandwidth $B$, operating at a carrier frequency $f_c$,
which in the context of this paper, 
is assumed to be below 6 GHz.
This infrastructure provides service to a set of \glspl{UE} $\mathcal K$, with $|\mathcal K|=K$,
arbitrarily distributed in the coverage area. 
Each \gls{UE} is equipped with $N_\text{UE}$ antenna elements and expects the reception of $N_\text{UE}^k$ independent data streams, with $N_\text{UE}^k\leq N_\text{UE}$ \footnote{
Note that this implicitly implies that the $k$th \gls{UE}  activates $N_\text{UE}^k\leq N_\text{UE}$ antennas.}. 
We stress at this point that determining adequate values for $N_\text{UE}^k$ constitutes one of the main contributions of this paper, 
and it is treated in Section V.A in detail.
To simplify forthcoming notation, 
we define $N_\text{str}=\sum_{k=1}^K N_\text{UE}^k$ as the total number of independent data streams throughout the network. 
Note that under the traditional \gls{M-MIMO} regime, 
it holds that $LSN_\text{TRP}\gg N_\text{str}$.
As also usually assumed in \gls{M-MIMO} contexts, 
the network relies on a \gls{TDD} transmission protocol,
whereby every \gls{TDD} frame matches the channel coherence interval (of size $\tau_c$, measured in samples or channel uses),
and it is divided into a UE-to-TRP pilot transmission phase, of size $\tau_p$, an uplink data transmission phase, of size $\tau_u$, and a downlink data transmission phase, of size $\tau_d$,
satisfying $\tau_c=\tau_p+\tau_u+\tau_d$.
For conciseness, in this paper, we will solely focus on the downlink performance, 
notwithstanding the fact that most insights also apply to the uplink segment.

\subsection{Channel Model}

The propagation and antenna models adopted follow the recommendations of the \gls{3GPP} Urban Macrocell model described in \cite{itu2017}. 
For simplicity, we assume that all \glspl{UE} experience \gls{NLOS} conditions. 
We will denote by $\beta_{mk}$ the large-scale gain  between the $m$th \gls{TRP} and the $k$th \gls{UE},
which is conformed by the product of three components, $\beta_{mk}=\iota_{mk}\chi_{mk}\Gamma_{mk}$, detailed next. 
The path loss $\iota_{mk}$ between the $m$th \gls{TRP} and the $k$th \gls{UE} is chosen in accordance to \cite{3GPP17}, 
but has been modified to consider the presence of indoor \glspl{UE} by adding a fixed wall penetration attenuation, i.e.,
\begin{equation}
    \iota_{mk}=\left\{
                    \begin{array}{cc}
                      \iota_0+10\alpha\log_{10}(d_{mk}) & \text{ outdoor UEs} \\
                      \iota_0+10\alpha\log_{10}(d_{mk})+\Delta_\text{indoor} & \text{ indoor UEs, } \\
                    \end{array}
                  \right.
\end{equation}
where $\iota_0$ is the path loss at the reference distance, 
$\alpha$ is the path loss exponent,
$\Delta_\text{indoor}$ is the wall penetration loss, 
and $d_{mk}$ is the distance between the $m$th \gls{TRP} and the $k$th \gls{UE}. 
For completeness, 
we also define $P_\text{ind}$ as the probability for an arbitrary \gls{UE} to be located indoors. The shadowing $\chi_{mk}$ is modeled as a correlated log-normal random variable with variance $\sigma_\chi^2$.
The model describing its spatial correlation can be found in \cite{ngo2017}.
The radiating patterns of the antenna elements forming the antenna array installed at each \gls{TRP} follow the specifications in \cite[Table 9]{itu2017}. 
In the sequel, we will use $\vartheta_{mk}$ to denote the nominal \gls{AoA} of the link between the $m$th \gls{TRP} and the $k$th \gls{UE}, 
seen from the \gls{TRP}, 
and $\Gamma_{mk}$ to represent the directional gain of the antenna array in the direction of the $m$th \gls{TRP} towards the $k$th \gls{UE}.

It is assumed that the multi-path is characterized by spatially correlated Rayleigh fading, 
where the resulting channel matrix can be defined as $\bs{G}_{mk}=\left[\bs g_{mk1} \hdots \bs g_{mkN_\text{UE}^k} \right] \in \mathbb{C}^{N_\text{TRP} \times N_\text{UE}^k}$,
with $\bs g_{mkn}$ denoting the channel vector from the $m$th \gls{TRP} to the $n$th antenna on the $k$th \gls{UE} (including both large-scale and small-scale fading). 
Spatial correlation matrices at both the transmit and receive antenna arrays are considered to conform to the popular Kronecker model, 
thus holding
$$\bs{G}_{mk}=(\bs R_{mk}^\text{TRP})^{1/2}\bs{G}_{mk}^\text{iid}(\bs R_{mk}^\text{UE})^{T/2},$$
with $\bs{G}_{mk}^\text{iid}$ representing an $ N_\text{TRP} \times N_\text{UE}^k$ matrix of \gls{iid} zero-mean circularly symmetric complex Gaussian random variables, 
and $\bs R_{mk}^\text{UE}$ and $\bs R_{mk}^\text{TRP}$ representing the spatial correlation matrices characterizing the scattering in the proximity of the \gls{UE} and \gls{TRP} antenna arrays, respectively. 
Following the model in \cite[Chapter 2]{bjornson2017}, 
$\bs R_{mk}^\text{TRP}$ and $\bs R_{mk}^\text{UE}$ can be calculated assuming an \gls{AoA} that follows a Gaussian distribution around the nominal angles $\vartheta_{mk}^\text{TRP}$ and $\vartheta_{mk}^\text{UE}$, respectively.
Defining $\bs{g}_{mk}=\text{vec}(\bs{G}_{mk})$, 
it is easy to check that
$$\bs{g}_{mk}\sim\mathcal C\mathcal N(\bs 0,\bs R_{mk}),$$
with $\bs R_{mk}=\bs R_{mk}^\text{UE}\otimes \bs R_{mk}^\text{TRP}$ and $\beta_{mk}=\frac{1}{N_\text{TRP}N_\text{UE}^k}\text{Tr}(\bs R_{mk})$.

Considering scenarios where \glspl{UE} move slowly,
it is reasonable to assume that the scattering covariance matrices $\bs R_{mk}^\text{TRP}$ and $\bs R_{mk}^\text{UE}$ change slowly,
and can be perfectly known at the \gls{CPU} and at the \glspl{TRP}\cite{ngo2017}.
For later use, 
we define now the $M N_\text{TRP} \times N_\text{UE}$ composite channel between the $k$th \gls{UE} and all the \glspl{TRP} as $\bs G_k=\left[\bs G_{1k}^T \hdots \bs G_{Mk}^T \right]^T$, 
and its vectorized version as $\bs g_k=\text{vec}(\bs G_k)$.  
The collective covariance matrix for the channel between the $k$th \gls{UE} and all the \glspl{TRP} can thus be defined as $\bs R_k=E\{\bs g_k\bs g_k^H \}=\mathcal D(\bs R_{1k}, \hdots, \bs R_{Mk})$.

\subsection{Uplink training and channel estimation}

In order to enable the estimation of the channel over each coherence time at the \glspl{TRP},
each \gls{UE} transmits a set of pilots, here referred to as pilot matrix $\bs \Phi_k$ of dimension $\tau_p \times N_\text{UE}^k$, 
that is, $\bs \Phi_k=[\bs\varphi_{k1} \hdots \bs\varphi_{kN_\text{UE}^k}]$, 
with $\bs\varphi_{kn}$ denoting the $\tau_p \times 1$ pilot sequence allocated to the $n$th antenna of the $k$th \gls{UE}. 

Interference among the antennas of \glspl{UE} ---or a given \gls{UE}--- during the channel estimation process is avoided by using a matrix of orthogonal pilot sequences, 
thus allowing the estimation of each individual channel $\bs g_{mkn}$. 
This implies that each pilot matrix $\bs \Phi_k$ fulfills $\bs \Phi_k^H \bs \Phi_k=\bs I_{N_\text{UE}^k}$. 
Whenever $N_\text{str}\leq \tau_p$, 
each \gls{UE} can be allocated a pilot matrix orthogonal to the ones assigned to other \glspl{UE},
i.e., $\bs \Phi_k^H \bs \Phi_{k'}=\bs 0_{N_\text{UE}^{k}\times N_\text{UE}^{k'}}$ for any $k,k'$. 
On the contrary, 
when $N_\text{str}> \tau_p$, 
some or all pilot matrices may need to be reused. 
In this work, the fingerprinting training introduced in \cite{femenias2019} is applied, 
whereby pilot sequences are reused only by \glspl{UE} which are located far apart from each other,
aiming at reducing the pilot contamination effects. 
Note that a desirable side effect of guaranteeing the orthogonality among the pilot sequences transmitted from the different antennas of a \gls{UE} is that channel estimation can be conducted independently for each antenna.

Denoting by $P^\text{UE}_p$ the available transit power per pilot at the \gls{UE},
the received training samples at the $m$th \gls{TRP} can be collected in an $N_\text{TRP}\times \tau_p$ matrix defined by
\begin{equation}
   {\bs{Y}_p}_m=\sum_{k=1}^K \sqrt{\tau_p P^\text{UE}_p/N_\text{UE}^k}\bs{G}_{mk} \bs{\Phi}_{k}^T+{\bs{N}_p}_m,
   \label{eq:pilotrecept}
\end{equation}
where ${\bs{N}_p}_m \in \mathbb{C}^{N_\text{TRP}\times\tau_p}$ is a matrix of \gls{iid} zero-mean circularly symmetric Gaussian random variables with standard deviation $\sigma_u$. 

An estimate of $\bs{G}_{mk}$ that minimizes the \gls{MMSE} can be derived by first projecting the received pilot matrix ${\bs{Y}_p}_m$ onto the \gls{UE}-specific pilot matrix \cite{mai2020}
\begin{equation}
\begin{split}
    \breve{\bs{Y}}_{p,mk}=&\left(\sum_{k'=1}^K \sqrt{\tau_p P^\text{UE}_p/N_\text{UE}^{k'}}\bs{G}_{mk'} \bs{\Phi}_{k'}^T\right)\bs{\Phi}_{k}^*\\
    &+{\bs{N}_p}_m\bs{\Phi}_{k}^*.
    \end{split}
\end{equation}

Using ${\bs g}_{mk}$ ---the vectorized form of the channel matrix, 
linking the $k$th \gls{UE} and the $m$th \gls{TRP}---,
it can be shown that the corresponding MMSE channel estimate follows \cite{kay1993,ngo2017,mai2020}
%\begin{equation}
%	\hat{\bs{g}}_{mkn}=\sqrt{\frac{\tau_p P_p^\text{UE}}{N_\text{UE}^k}} \bs{R}_{mk}^\text{TRP}\bs{\Psi}_{mk}^{-1} \breve{\bs{y}}_{p,mkn},
%\end{equation}
%where
%\begin{equation}
%   \bs{\Psi}_{mk}=\sum_{k'=1,k'\neq k}^K \sum_{n=1}^{N_\text{UE}^{k'}} \frac{\tau_p P_p^\text{UE}}{N_\text{UE}^{k'}}\bs{R}_{mk'}^\text{TRP} \left|\bs{\varphi}_{k'n}^H{\bs{\varphi}}_{kn}\right|^2+\sigma_{\upsilon}^2\bs{I}_{N_\text{TRP}}.
%\end{equation}
\begin{equation}
	\hat{\bs{g}}_{mk}=\sqrt{\frac{\tau_p P_p^\text{UE}}{N_\text{UE}^k}} \bs{R}_{mk}\bs{\Psi}_{mk}^{-1} \breve{\bs{y}}_{p,mk},
     \label{eq:channelest}
\end{equation}
where $\breve{\bs{y}}_{p,mk}=\text{vec}(\breve{\bs{Y}}_{p,mk})$ and
\begin{equation}
\begin{split}
   \bs{\Psi}_{mk}=&\sum_{k'=1}^K \frac{\tau_p P_p^\text{UE}}{N_\text{UE}^{k'}}
   (\bs{\Phi}_{k}^H\bs{\Phi}_{k'}\otimes\bs I_{N_\text{TRP}})\bs{R}_{mk'}(\bs{\Phi}_{k'}^H\bs{\Phi}_{k}\otimes\bs I_{N_\text{TRP}})\\
   &+\sigma_{\upsilon}^2\bs{I}_{N_\text{TRP}N_\text{UE}^k}.
   \end{split}
\end{equation}

Moreover, the distribution of the MMSE channel estimate can be shown to be 
$$\hat{\bs g}_{mk}\sim \mathcal C\mathcal N\left(\bs 0,\frac{P_p^\text{UE}\tau_p}{N_\text{UE}^k}\bs{R}_{mk}\bs{\Psi}_{mk}^{-1}\bs{R}_{mk}\right),$$ 
whereas the distribution of the channel estimation error $\tilde{\bs g}_{mk}=\bs{g}_{mk}-\hat{\bs{g}}_{mk}$ conforms to $\tilde{\bs g}_{mk}\sim \mathcal C \mathcal N(\bs 0,\bs A_{mk})$, 
with $$\bs A_{mk}=\bs{R}_{mk}-\frac{P_p^\text{UE}\tau_p}{N_\text{UE}^k}\bs{R}_{mk}\bs{\Psi}_{mk}^{-1}\bs{R}_{mk}.$$
Relying on the orthogonality of $\bs{\Phi}_k$, 
it is easy to check that the error covariance matrix possesses a block-diagonal structure, 
i.e. $$\bs A_{mk}=\mathcal D(\bs A_{mk1},\hdots, \bs A_{mkN_\text{UE}^k}).$$
For convenience, we implicitly define $\hat{\bs G}_{mk}$, $\hat{\bs G}_k$ and $\hat{\bs g}_k$ as the MMSE-estimated counterparts of ${\bs G}_{mk}$, ${\bs G}_k$ and $\bs g_k$, respectively. 
It is worth pointing out at this stage that if a \gls{UE} $k'$ only activates $N_\text{UE}^{k'}<N_\text{UE}$ antennas 
(i.e., it plans the reception of $N_\text{UE}^{k'}$ data streams), 
its corresponding matrix pilot $\bs \phi_{k'}$ only has dimensions $\tau_p \times N_\text{UE}^{k'}$ with $\bs \Phi_{k'}^H \bs \Phi_{k'}=\bs I_{N_\text{UE}^{k'}}$.

\section{Baseband processing}
\label{sec:BB}

In this section, 
we introduce the baseband protocols used in this paper. 

\subsection{\gls{UE} association and \gls{TRP} (de)activation}
In the pursue of \gls{EE},
we  embrace that a number of \glspl{TRP} can be shut down,
and thus consider that only $M\leq M_T $ \glspl{TRP} are active at any given instant.
Following a \gls{UE}-centric paradigm (see \cite{demir2021} for details),
where each \gls{UE} is served only by a subset of the active \glspl{TRP},
we denote by $\bs C(\tau)$ the $M \times K$ connectivity matrix indicating which \glspl{TRP} serve each \gls{UE} over a specific large-scale propagation interval $\tau$.
Its  entries $c_{mk}(\tau)$ are configured as
\begin{equation}
    c_{mk}(\tau)=\left\{\begin{array}{cc}
          1 & \text{UE } k \text{ served by TRP } m \text{ over interval } \tau \\
          0 & \text{otherwise.}
        \end{array}\right.
        \label{connect}
\end{equation}

For convenience,
and since results in the following subsections are derived for a single arbitrary large-scale interval,
the index $\tau$ will be dropped from the connectivity matrix,
i.e., $\bs C$ and $c_{mk}$ will be used instead of $\bs C(\tau)$ and $c_{mk}(\tau)$, respectively.
We also define at this point the $1 \times K$ vector $\bs c_{[m]}$ as the $m$th row of $\bs C$,
which represents the connectivity of the $m$th \gls{TRP},
and the $M\times 1$ vector $\bs c^{[k]}$ as the $k$th column of $\bs C$,
which corresponds to the connectivity of the $k$th \gls{UE}.
We also denote sets $\mathcal M_T=\left\{m_T^1, \cdots,  m_T^{M_T}\right\}$ and $\mathcal M=\left\{m^1, \cdots,  m^{M}\right\}$ as the collections of all \glspl{TRP} in the network and the active ones, respectively,
holding $\mathcal M \subseteq \mathcal M_T$.
Section \ref{sec:adaptive} discusses our new method to devise the \glspl{TRP} to shut down,
filling in the entries of \eqref{connect}.

\subsection{Transmitter processing}
The samples received by the $k$th \gls{UE} can be expressed as
\begin{equation}
    \bs y_{k}=\bs G_k^T \tilde{\bs W}\bs s+ \bs\upsilon_{k},
\label{eq:recept}
\end{equation}
where 
$\bs s=\left[\bs s_1^T \hdots s_K^T \right]^T$ is the $N_\text{str}\times 1$ vector of transmitted symbols, 
with $E\{\bs s^H\bs s\}=\bs I_{N_\text{str}}$ and $\bs s_k=[s_{k1} \hdots s_{kN_\text{UE}^k}]^T$, 
with $s_{kn}$ representing the information symbol for the $k$th \gls{UE} on the $n$th stream,
$\tilde{\bs W}$ is the $MN_\text{TRP}\times N_\text{str}$ precoding matrix (including transmit power allocation),
and $\bs \upsilon_{k}\sim \mathcal {CN}(\bs 0,\sigma_\upsilon^2 \bs I_{N_\text{UE}^k})$ is a vector of \gls{AWGN} samples.

For convenience, 
we note that the global precoding matrix $\tilde{\bs W}$ can be defined on the basis of the TRP-specific blocks, that is, $\tilde{\boldsymbol W}=\left[\tilde{\boldsymbol W}^1 \hdots \tilde{\boldsymbol W}^M\right]^T$,
where $\tilde{\boldsymbol W}^m$ is the $N_\text{TRP}\times N_\text{str}$ precoding matrix applied at the $m$th \gls{TRP}. 
It fulfils the transmit power constraint, $\|\tilde{\boldsymbol W}^m\|_F^2\leq P_{TRP} \ \forall m$.
Alternatively, 
the precoder can be examined from a \gls{UE} perspective by partitioning $\tilde{\bs W}$ as $\tilde{\bs W}=\left[\tilde{\bs W}_1 \hdots \tilde{\bs W}_K \right]$, 
where $\tilde{\bs W}_k$ is the $MN_\text{TRP}\times N_\text{UE}^k$ block affecting the symbols $\bs s_k$ transmitted towards the $k$th \gls{UE}. 
In turn, $\tilde{\bs W}_k$ can be decomposed into its constituent stream-oriented precoders as
$\tilde{\boldsymbol W}_k=\left[\tilde{\bs w}_{k1} \hdots \tilde{\bs w}_{kN_\text{UE}^k}\right]$, 
allowing the received samples in \eqref{eq:recept} to be rewritten as
\begin{equation}
\begin{split}
  \bs y_k&=\bs G_k^T\sum_{k'=1}^{K}\tilde{\bs W}_k \bs s_k + \bs\upsilon_k \\
         &= \bs G_{k}^T\sum_{k'=1}^K \sum_{n=1}^{N_\text{UE}^{k'}}\tilde{\bs w}_{k'n}s_{k'n}+\bs \upsilon_{k}.
\end{split}
\label{eq:recept2}
\end{equation}

In this paper, 
and without loss of generality, 
we use the \gls{CP-MMSE} precoder design introduced in \cite{bjornson2019}. 
This precoder strives at a compromise between the strength of the desired signal transmitted at each \gls{UE} and the interference it generates among the different \glspl{UE}, 
wile preserving the scalability of the network. %The \gls{CP-MMSE} precoder stems from the full \gls{MMSE} precoder, which can be considered a performance upper bound due to its near optimality on the basis of the uplink-downlink duality theorem \cite{demir2021}.
As a result, the  $MN_\text{TRP}\times 1$ precoding vector targeting the $n$th stream of the $k$th \gls{UE} can be expressed as
\begin{equation}
     \tilde{\bs w}_{kn}=\sqrt{\frac{p^\text{DL}_{k}}{N_\text{UE}^k}}\frac{{\bs w}_{kn}}{\sqrt{E\{\|{\bs w}_{kn}\|^2 \}}},
     \label{cent_norm}
\end{equation}
where the downlink transmit power $p_{k}^\text{DL}$ allocated to the $k$th \gls{UE} is equally divided among its data streams,
and
\begin{equation}
\boldsymbol w_{kn}=\frac{p_k^{\text{UL}}}{N_\text{UE}^k}\boldsymbol{\Omega}_k^{-1}\boldsymbol{C}^{[k]}\hat{\boldsymbol{g}}_{kn},
\label{CMMSE}
\end{equation}
with
\begin{equation}
\begin{split}
\boldsymbol{\Omega}_k &=\sum_{i\setminus \bs {c^{[k]}}^T \bs c^{[i]}>0} \frac{p_i^\text{UL}}{N_\text{UE}^k}\left( \bs C^{[k]}(\hat{\bs G}_{i} \hat{\bs G}_{i}^H)\bs C^{[k]}+\bs C^{[k]} \bs A_{i} \bs C^{[k]}\right)\\
&+\sigma_\upsilon^2\bs I_{MN_\text{TRP}N_\text{UE}^k},
\end{split}
\end{equation}
%$$\bs Z_k=\sum_{i\setminus \bs {c^{[k]}}^T \bs c^{[i]}>0} p_i^\text{UL} \bs C^{[k]} \bs A_{i} \bs C^{[k]},$$
%and
%$$\bs D_{ki}=\bs C^{[k]}(\hat{\bs g}_{i} \hat{\bs g}_{i}^H+\bs A_{i})\bs C^{[k]},$$
where
$\bs C^{[k]}=\mathcal D([\overbrace{\bs c^{[k],T} \hdots \bs c^{[k],T}}^{N_\text{TRP}N_\text{UE}^k}])^T$
and
$\bs A_{i}=\mathcal D(\bs A_{1i} \cdots \bs A_{Mi})$. 
Note that the normalization step in \eqref{cent_norm} ensures that $E\{\|\tilde{\bs w}_{kn}\|^2\}=p_{k}^\text{DL}/N_\text{UE}^k$.

For the sake of preserving scalability principles,
fractional power allocation is adopted \cite{demir2021}, 
i.e.,
\begin{equation}
    p_{k}^\text{DL}=\frac{P_\text{TRP}(\sum_{m=1}^M c_{mk} \beta_{mk})^\upsilon}{\max_{l\in\{1,\hdots,M\}}c_{lk}\sum_{i=1}^{K}c_{li}(\sum_{m=1}^M c_{mi}\beta_{mi})^\upsilon},
    \label{power_dl}
\end{equation}
with $\upsilon \in [-1,1]$ denoting a parameter used to approximate the power allocation to different performance targets (e.g., sum-rate, max-min).
Also note that the precoder depends on the uplink power allocation policy,
which is set here in accordance to the uplink fractional power allocation described in \cite[eq. (7.34)]{demir2021}.
%\begin{equation}
%p_k^\text{UL}=P_\text{MS}\frac{(\sum_{m=1}^M c_{mk} \beta_{mk})^\upsilon}{\max_{i\in\{1,\hdots,{K^\text{CF}}\}}(\sum_{m=1}^M c_{mi}\beta_{mi})^\upsilon},
%\end{equation}
%with $P_\text{MS}$ denoting the available transmit power at each \gls{MS}.

\subsection{Receiver processing}
Having defined the processing done at the transmitter side, 
let us now consider the processing carried out at the \gls{UE}. 
Towards this end, 
let us define ${\bs \Xi}_{kk'}=\bs G_k^T \tilde {\bs W}_{k'}$, 
thus leading to yet another reformulation of the received samples,
i.e.,
\begin{equation}
\begin{split}
\bs y_k&=\sum_{k'=1}^{K}{\bs \Xi}_{kk'}\bs s_{k'} +\bs\upsilon_k\\
       &=\underbrace{{\bs \Xi}_{kk}\bs s_{k}}_{\text{Useful signal}}+\underbrace{\sum_{k'=1,k'\neq k}^{K}{\bs \Xi}_{kk'}\bs s_{k'}}_\text{Interference}+\underbrace{\bs\upsilon_k}_\text{Noise}.
\end{split}
\end{equation}
Under the assumption of statistical \gls{CSI} at the receiver (i.e., no pilots are transmitted on the downlink), 
only the mean of the equivalent channel affecting the useful term is available at the \gls{UE}.
Therefore it holds
\begin{equation}
\bs y_k=\bar {\bs \Xi}_k \bs s_k + \bs e_k,
\label{recept2}
\end{equation}
with  $\bar{\bs \Xi}_{k}=E\{{\bs \Xi}_{kk}\}$ representing the average equivalent channel affecting the $k$th \gls{UE}, 
and
\begin{equation}
\bs e_k=\sum_{k'=1}^{K}{\bs \Xi}_{kk'}\bs s_{k'} - \bar {\bs \Xi}_k \bs s_k + \bs \upsilon_{k}
\end{equation}
collecting the inter-\gls{UE} interference, the self-interference (due to reliance on statistical \gls{CSI}) and the \gls{AWGN} noise.

Two different reception strategies are considered ---namely, a linear one (\gls{MMSE}) and a non-linear one (\gls{MMSE}-\gls{SIC})---, 
which are based on the availability of statistical \gls{CSI}. 
These are now described in detail:
\begin{enumerate}
\item 
\gls{MMSE}: 
    A linear combiner is applied to the received samples to estimate the payload data symbols
    \begin{equation}
        \hat{\bs s}_k=\bs U_k^H \bs y_k=\bs U_k^H\bar {\bs \Xi}_k \bs s_k + \bs U_k^H \bs e_k,
    \end{equation}
    where re-expressing $\bs U_k=[\bs u_{k1} \hdots \bs u_{kN_\text{UE}^k}]$ and $\bar {\bs \Xi}_k=[\bar {\bs \xi}_{k1} \hdots \bar {\bs \xi}_{kN_\text{UE}^k}]$ allows the estimate of the $n$th symbol for the $k$th \gls{UE} to be written as
    \begin{equation}
        \hat s_{kn}=\bs u_{kn}^H\bar{\bs \xi}_{kn} s_{kn}+\bs u_{kn}^H\bs b_{kn},
        \label{estimated}
    \end{equation}
    with $\bs b_{kn}=\sum_{n'=1,n'\neq n}^{N_\text{UE}^k}\bar{\bs \xi}_{kn'}s_{kn'}+\bs e_k$.
    An achievable downlink \gls{SE} in bit/s/Hz for the $k$th \gls{UE} can now be obtained as \cite{demir2021}
    \begin{equation}
        R_k=\frac{1-\tau_p/\tau_c}{2}\sum_{n=1}^{N_\text{UE}^k}\log_2(1+\text{SINR}_{kn}),
        \label{eq:rate}
    \end{equation}
    noting that the 1/2 factor is due to the assumption of an equal UL-DL split,
    and on the basis of \eqref{estimated}, 
    the instantaneous effective SINR for the $n$th stream of the $k$th \gls{UE} follows
    \begin{equation}
        \text{SINR}_{kn}=\frac{|\bs u_{kn}^H\bar{\bs \xi}_{kn}|^2}{\bs u_{kn}^H \bs B_{kn}\bs u_{kn}},
        \label{SINR}
    \end{equation}
    where
    \begin{equation}
        \bs B_{kn}=E\left\{\bs b_{kn} \bs b_{kn}^H \right\}=\sum_{n'=1,n'\neq n}^{N_\text{UE}^k}\bar{\bs \xi}_{kn'}\bar{\bs \xi}_{kn'}^H+\bs E_k,
    \end{equation}
    with
    \begin{equation}
        \begin{split}
        \bs E_k=E\left\{\bs e_k\bs e_k^H\right\}=&E\left\{ \bs G_k^T \tilde{\bs W}\tilde{\bs W}^H\bs G_k^*\right\}\\
        &-\bar{\bs \Xi}_k \bar{\bs \Xi}_k^H+\sigma_\upsilon^2 \bs I_{N_\text{UE}^k}.
        \end{split}
    \end{equation}
    Since \eqref{SINR} has the form of a generalized Rayleigh quotient, 
    this can be maximized by resorting to the \gls{MMSE} combiner defined by \cite{demir2021}
    \begin{equation}
      \bs u_{kn}=\bs B_{kn}^{-1}\bar {\bs \xi}_{kn},
    \end{equation}
    thus resulting in a maximum achievable \gls{SE}
    \begin{equation}
        R_k^\text{MMSE}=\frac{1-\frac{\tau_p}{\tau_c}}{2}\sum_{n=1}^{N_\text{UE}^k}\log_2(1+\bar {\bs \xi}_{kn}^H\bs B_{kn}^{-1}\bar {\bs \xi}_{kn}),
        \label{eq:finalrate}
    \end{equation}

\item 
\gls{MMSE}-\gls{SIC}: 
    \gls{SIC} has been successfully adapted to the downlink of a cell-free network under the assumption of statistical \gls{CSI} at the receiver \cite{mai2020} resulting in a \gls{SE}
    \begin{equation}
        R_k^\text{MMSE-SIC}=\frac{1-\frac{\tau_p}{\tau_c}}{2}\log_2\left|\bs I_{N_\text{UE}^k}+\bar{\bs \Xi}_k^H \bs E_k^{-1} \bar{\bs \Xi}_k\right|.
        \label{eq:finalrateSIC}
    \end{equation} 
\end{enumerate}

%Anticipating our interest in the \gls{EE} aspects of the considered network,
%it is convenient to have an \gls{TRP}-oriented view of the precoder $\tilde{\boldsymbol W}$.

\section{Power consumption model}

To model the network power consumption,
we use the realistic 4G/5G \gls{M-MIMO} \glspl{AAU} model presented in \cite{piovesan2022}.
In particular,
and restricting to the case where each \gls{AAU} operates a single carrier/cell,
the power consumed by the \gls{AAU} of the $m$th \gls{TRP} when actively serving \glspl{UE} is given by
\begin{equation}
    \tilde P^\text{AAU}_m=\underbrace{P^\text{AAU,fix}+P^\text{AAU,BB}+P_{TRX}+P_{PA}}_{P_{Bline}}
            +\underbrace{\frac{1}{\eta}P^{TX}_m}_{P_{out}}
    \label{eq:powermodel}
\end{equation}
where
$P^\text{AAU,fix}$ accounts for part of the \gls{AAU} circuitry that is always active (e.g., circuitry used to control the \gls{AAU} activation/deactivation),
$P^\text{AAU,BB}$ is the power required for the baseband processing performed at the \gls{AAU},
$P_{TRX}=\sum_{t=1}^T M_{av,t}D_{TRX,t}$ is the power consumed by the $T$ transceivers in the \gls{AAU}
(which can be calculated as the product of the number, $M_{av,t}$, of available RF chains and the power, $D_{TRX,t}$, consumed by each RF chain),
$P_{PA}$ is the static power consumed by the power amplifiers (PAs),
and $P_{out}$ is the power consumed by the transmit power
(which is equal to the ratio of the transmit power, $P^{TX}_m=E\{\|\tilde{\boldsymbol W}^m\|_F^2\}$, to the efficiency of the PAs and antennas $\eta$).
Note that the transmit power usually increases linearly with the number of resources utilized,
and also depends on the specific baseband precoder implemented by each \gls{TRP}.
The first four terms are referred to as the baseline power consumption,
denoted by $P_{Bline}$,
and represent the overall power that an \gls{AAU} requires to operate when there is no load.

As pointed out in \cite{lopez2022},
the ultra-lean signalling design,
first proposed in \gls{3GPP}  \gls{NR},
paves the way for a more energy efficient operation of the \glspl{TRP}. In particular,
symbol shutdown mechanisms are enabled that can be operated over short timescales (i.e., from \gls{OFDM} symbol level to 160\,ms),
and that can be combined with carrier shutdown and deep dormancy strategies,
which progressively switch off more \gls{AAU}  hardware
but can only react over coarser time scales (i.e., from seconds to minutes the former, and minutes and hours the latter).

Considering these shutdown methods,
formally,
the power consumption of the $m$th \gls{TRP} can now be rewritten as
\begin{equation}
    P^\text{AAU}_{m}=\left\{\begin{array}{cc}
                       \tilde P^\text{AAU}_m & \text{ if } m \in \mathcal M \\
                       (1-\varpi)P_{Bline} & \text{ if } m \in \mathcal M_T \setminus \mathcal M,
                     \end{array}
                     \right.
    \label{eq:powermodelAAU}
\end{equation}
where $\varpi$ is the fraction of the baseline power saved when a \gls{TRP} is in a shutdown mode.
Recent studies have shown that, statistically, $\varpi$ takes average values of 0.3, 0.47 and 0.7 when using symbol shutdown, carrier shutdown and dormancy, respectively \cite{piovesan2022}.

\begin{figure}
      \begin{center}
       \includegraphics[width=\linewidth]{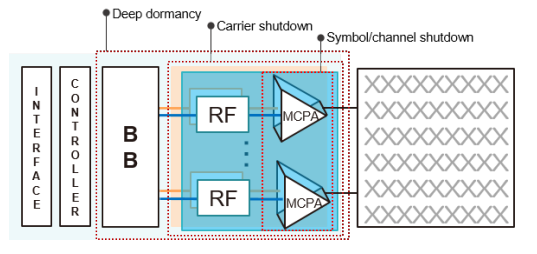}
      \end{center}
      \caption{Power consumption model for \gls{AAU} \cite{piovesan2022}.}% Circled \gls{UE} depicts a \gls{UE} that has been offloaded from the terrestrial to the satellite segment.}
      \label{fig:powermodel}
\end{figure}

Aside from the \gls{AAU} power consumption,
and unlike classical cellular topologies,
the \gls{CF-M-MIMO} scheme must also consider the power consumption related to the extra infrastructure it requires,
namely, the fronthaul links and the \gls{CPU}.
Following the results in \cite{chen2022},
the power consumption of the fronthaul linking the $m$th \gls{TRP} to the \gls{CPU} over the downlink transmission phase can be expressed as
\begin{equation}
    P_m^\text{FH}=\left\{\begin{array}{cc}
                        P_m^\text{FH,fix}+\frac{\tau_d}{\tau_c}K_\text{TRP}P_m^\text{FH,var} & \text{ if } m \in \mathcal M \\
                        P_m^\text{FH,fix}& \text{ if } m \in \mathcal M_T \setminus \mathcal M,
                     \end{array}
                     \right.
    \label{eq:powermodelFH}
\end{equation}
where $P_m^\text{FH,fix}$ and $P_m^\text{FH,var}$ are the fixed and rate-dependent power consumption of the $m$th fronthaul, respectively.
The power consumption of the \gls{CPU} can be modelled as \cite{chen2022}
\begin{equation}
    P^\text{CPU}=P^\text{CPU,fix}+B\sum_{k=1}^{K}R_k P^\text{CPU,pre},
    \label{eq:powermodelCPU}
\end{equation}
where $P^\text{CPU,fix}$ and $P^\text{CPU,pre}$ are the fixed and precoding-dependent power consumption of the \gls{CPU}, respectively
---the latter one being expressed in W/Gbps.

Having established the different terms,
the total power consumption for the whole network when considering the full set $\mathcal M_T$ of \glspl{TRP} is given by
\begin{equation}
    P^\text{TOT}=\sum_{m=1}^{M_T} \left(P^\text{AAU}_{m}+ P_m^\text{FH}\right)+P^\text{CPU}.
    \label{eq:powertotal}
\end{equation}

Relying on \eqref{eq:rate} and \eqref{eq:powertotal},
we can now define the energy efficiency, $EE$, of the whole network over an arbitrary large-scale time interval as
\begin{equation}
    EE=\frac{B\sum_{k=1}^K R_k}{P^\text{TOT}} [bits/J].
    \label{eq:EE}
\end{equation}

\section{Sleep mode and stream management}
Having established the main processing steps and energy consumption requirements of a \gls{CF-M-MIMO} network, and before delving into adaptive strategies, it is important to recap and clarify the time scales that apply to the different operations. In particular, three different time scales are considered here:
\begin{enumerate}
    \item \emph{Short-scale}: It is basically defined by the coherence time of the wireless channel. Pilot transmission (eq.~\eqref{eq:pilotrecept}), channel estimation (eq.~\eqref{eq:channelest}) and precoder design (eq.~\eqref{cent_norm}) are all processing steps that are conducted over a short-time scale. Power adaptation at this time scale is related to symbol shutdown mechanisms.
    \item \emph{Large-scale}: This is characterized by the changes experimented by large-scale propagation conditions (e.g., path loss, shadowing loss) and it is usually considered to span 40 to 100 channel coherence times. Pilot and power allocation (eq.~\eqref{power_dl}) are most often conducted on large-scale time basis as well as the combiner design at the receiver when relying only on statistical channel information. Power adaptation at this time scale is related to carrier shutdown mechanisms.
    \item \emph{\gls{UE} dynamics scale}: This time scale is related to \gls{UE} dynamics whose mobility patterns can span from minutes to hours or even days and weeks. It serves to define \gls{UE} spatial probabilities that govern the \glspl{UE} distribution across the coverage area. Power adaptation at this time scale is exploited using the \gls{TRP} dormancy state.
\end{enumerate}
In the coming subsections, we focus on adaptive mechanisms that are typically conducted at a large-scale (carrier shutdown), basically, because they exploit the large-scale propagation parameters (e.g., large-scale channel gains, correlation matrices). Particularly, in the two mechanisms covered next, \gls{UE} stream management and \gls{TRP} selective (de)activation, they are executed whenever significant changes in the large-scale propagation conditions take place with the decisions taken being effective over many coherence times. Nonetheless, some results are provided to hint what benefits
bring the different power shutdown modes.

\subsection{Adaptive UE stream management}
\label{sec:adapt_stream}

As described in Sec.~\ref{sec:model}, 
it is assumed that all \glspl{UE} are equipped with $N_\text{UE}$ antennas, 
thus allowing the reception of up to $N_\text{UE}$ data streams. 
However, as it will be shown next, 
it is not always advisable to allow the maximum degree of multiplexing to all \glspl{UE}. 
The \glspl{UE} \gls{ASM} described here refers to how to select the number of data streams ($N_\text{UE}^k$) per \gls{UE} to optimize network performance.
To this end, 
it is critical to recognize that allowing \glspl{UE} to receive more than one data stream does not necessarily lead to enhanced performance due to two fundamental reasons:
\begin{enumerate}
\item 
Power split: 
All practical transmit power allocation strategies are rooted on large scale parameters (i.e., covariance matrices).
Hence, the power devoted to a given \gls{UE} on a per-coherence time basis (i.e., instantaneously) can only be uniformly split among the transmitted streams. 
This power division can cause that \glspl{UE} experiencing bad propagation conditions have their \gls{SINR} further degraded \cite{mai2020}.
\item 
Increased pilot contamination: 
Antennas on any given \gls{UE} must be assigned orthogonal pilots in order for their channels to be separated upon estimation,
thus the constraint $\bs \Phi_k^H \bs \Phi_k=\bs I_{N_\text{UE}^k}$. 
Given a pilot length $\tau_p$, 
the pilot reuse factor, 
which in turn defines the number of interferers affecting the estimation of each channel, 
will be given by\footnote{
Since pilot allocation is based on large-scale parameters, 
this expectation operator is used to average across many realizations of random \gls{UE} placements.} 
$\frac{\tau_p}{E\{N_\text{str}\}}$, 
which reduces to $\tau_p/K$ when all \glspl{UE} just receive one single data stream and to $\tau_p/(K N_\text{UE})$ when all \glspl{UE} aim at receiving as many data streams as antennas they have \cite{li2016}. 
Logically, the more antennas the \glspl{UE} activate, 
the larger the pilot contamination effect becomes.
\end{enumerate}

\begin{figure}[!t]
\vspace{0.1cm}
      \begin{center}
       \includegraphics[width=\linewidth]{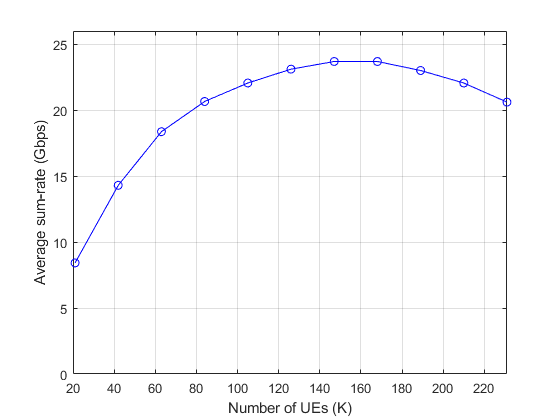}
      \end{center}
      \caption{CF-M-MIMO sum-rate under different network configurations ($M=21$ \glspl{TRP}, $\tau_p=30$ samples, $N_\text{TRP}=64$ antennas, $N_\text{UE}=1$ antenna).}
      \label{sumratediffK}
\end{figure}

Before delving into strategies on how to manage the \glspl{UE}' multiple antennas, 
it is worth assessing the performance of a \gls{CF-M-MIMO} network when increasing the load for the specific case of single-antenna \glspl{UE}. 
In particular, we focus on the total network rate (sum-rate) given by $R=\sum_{k=1}^K R_k$, 
where $R_k$ follows from \eqref{eq:finalrate}, 
given a specific network configuration defined by the main setup parameters, $\bs \Omega=\{M, d_\text{ISD}, N_\text{TRP}, \tau_p \}$, 
where $d_\text{ISD}$ denotes the inter-site distance among neighbouring \glspl{BS}. 
Figure \ref{sumratediffK} depicts the sum-rate when considering a network following the topology in Fig.~\ref{topology},
and when increasing the number of \glspl{UE} per cell one \gls{UE} per cell at a time 
(i.e., $K=21, 42, 63, \hdots$).
It can be observed how the overall network sum-rate first increases\footnote{
Results should be generated for any arbitrary system configuration $\bs \Omega'$.} 
as more \glspl{UE} are admitted into the system, 
up to a point where it starts to decline, 
thus revealing that the pilot contamination effects outweigh the multi-\gls{UE} diversity plus multiplexing gains. 
For the situation at hand, $\bs \Omega=\left\{21,200,64,30\right\}$, 
the optimum number of \glspl{UE} seems to be approximately $K_{\bs \Omega}^\text{opt}\backsimeq150$ (roughly 7 single-antenna \glspl{UE} per hexagon in Fig.~\ref{topology}). 
%The following observation is key to develop an effective stream handling strategy:\\
%\textbf{Result:}El millor que pot passar es que cada stream independent vengui vagi a un usuari diferent\\

In the following, we propose a heuristic strategy to decide on the number of data streams per \gls{UE} that, in alignment with the general cell-free \emph{philosophy}, 
aims at protecting the worst-performing \glspl{UE} in the network at the cost of sacrificing some of the potential improvements in rate for the \glspl{UE} that are better off. 
Nonetheless, this strategy is generic enough to be adapted to other objectives. 

The scheme starts by determining the maximum number of streams for a specific network layout $N_\text{str}$, 
and then fixing the groups of \glspl{UE} with a prescribed number of streams, denoted as $\mathcal K^{(1)}, \hdots,\mathcal  K^{(N_\text{UE})}$, 
with $\mathcal K^{(i)}$ representing the set of \glspl{UE} receiving $i$ data streams,
whose cardinality is $K^{(i)}$.
Note that $N_\text{str}\leq K_{\bs \Omega}^\text{opt}$ and that $\sum_{i=1}^{N_\text{UE}} K^{(i)}=K$.  
Towards this end, 
each \gls{UE} $k$ is then assessed on the basis of two metrics:
\begin{enumerate}
  \item Conditioning: 
  $\xi_k=\lambda_k^\text{max}/\lambda_k^\text{min}$, 
  with $\lambda_k^\text{max}$ and $\lambda_k^\text{min}$ denoting the maximum and minimum eigenvalues of ${\bs R}_k$. 
  This metric measures the suitability of the corresponding channels to be favourably inverted by the \gls{MMSE} filter design in \eqref{CMMSE}.
  \item Strength: 
  ${\beta}_k=\text{Trace}({\bs R}_k)$. 
  This metric relates to the overall propagation gain a given \gls{UE} is subject to.
\end{enumerate}
For simplicity let us assume that \glspl{UE} are split into two groups: 
\emph{Good} \glspl{UE} are those experiencing good conditioning and large strength, 
that is, those \glspl{UE} with $\xi_k<\theta_{\xi}^{\text{good}}$ and ${\beta}_k>\theta_{\beta}^{\text{good}}$, 
with $\theta_{\xi}^{\text{good}}$ and $\theta_{\beta}^{\text{good}}$ representing adequate thresholds that must be determined by simulation. 
In the numerical results section, 
specific numerology is given, 
but it has been found that overall network performance is not overly sensitive to these values. 
These \glspl{UE}  are the ones that can safely support multi-stream transmission, 
and thus potentially conform sets $\mathcal K^{(2)}, \hdots,\mathcal  K^{(N_\text{UE})}$. 
In contrast, 
\emph{bad} \glspl{UE} are those subject to bad conditioning and large propagation losses, 
and thus must be limited to single-stream transmissions, 
and assigned to $\mathcal K^{(1)}$. 
In general setups, \glspl{UE} will need to be classified into $N_\text{UE}$ groups, 
thus the quantitative evaluation of \emph{goodness} should be done by fixing $N_\text{UE}-1$ thresholds on both metrics ($\theta_{\xi}^{\text{good},1},\hdots, \theta_{\xi}^{\text{good},N_\text{UE}-1},\theta_{\beta}^{\text{good},1},\hdots, \theta_{\beta}^{\text{good},N_\text{UE}-1}$). 
To this end, 
the following recursive strategy is proposed here: 
We start by aiming at a first split of \glspl{UE} into two groups, $\mathcal K^{(1)}$ and $\mathcal K_{\text{rest}}^{(1)}$, formed by those \glspl{UE} that are assigned one and more than one data streams, respectively. 
The objective now is to guarantee that \glspl{UE} in $\mathcal K^{(1)}$ achieve virtually the same performance as if $\mathcal K^{(1)}=\mathcal K$ 
(all \glspl{UE} were assigned one data stream). 
As a result, 
the set of all available pilot sequences, $\mathcal P$, is split into two groups $\mathcal P_1$ and $\mathcal{P}_{\text{rest}}^{(1)}$, 
where it should be guaranteed that $$\frac{K^{(1)}}{|\mathcal P_1|}=\frac{K}{\tau_p},$$
hence
\begin{equation}
    |\mathcal P_1|=\text{round}\left\{\frac{K^{(1)}\tau_p}{K}\right\} \ \text{and} \ |\mathcal{P}_{\text{rest}}^{(1)}|=\tau_p-|\mathcal P_1|.
    \label{pilotpartition}
\end{equation}
The split of pilots according to \eqref{pilotpartition} aims at guaranteeing that the worst \glspl{UE} are not affected by the increased pilot contamination due to multi-stream transmission. In fact, note that \eqref{pilotpartition} ensures that that \glspl{UE} in $\mathcal K^{(1)}$ have available (approximately), the same number of pilots they had in the initial setup when all \glspl{UE} had only one stream assigned. This procedure is generalized to as many groups as potential number of data streams by successively applying the same technique on the \gls{UE} and pilot sets, $\mathcal K_{\text{rest}}^{(1)}$ and $\mathcal P_\text{rest}$, respectively,
that is, splitting  $\mathcal K_{\text{rest}}^{(1)}$ into $\mathcal K^{(2)}$ and $\mathcal K_{\text{rest}}^{(2)}$, and $\mathcal{P}_{\text{rest}}^{(1)}$ into $\mathcal P_2$ and $\mathcal{P}_{\text{rest}}^{(2)}$. The general procedure to generate the \gls{UE} and pilot groups is detailed in Algorithm 1. As shown in the algorithmic description, and after suitably initializing all variables in use, the aim is to create up to a maximum of $N_\text{UE}$ groups of \glspl{UE} each characterized by the number of streams each group is able to handle (from $1$ up to $N_\text{UE}$). Towards this end, it first tries to determine whether there are \glspl{UE} with  good enough propagation conditions (as determined by $\theta_{\xi}^{\text{good,g}}$ and $\theta_{\beta}^{\text{good,g}}$) and subsequently evaluates whether there are enough pilot symbols to guarantee the contention of pilot contamination effects affecting the \glspl{UE} with worst channels. When both conditions are fulfilled, a new group $\mathcal K^{(g)}$ is born with a prescribed group-specific pilot allocation $\mathcal P_g$. If at any point in the loop the number of allocated streams $L$ exceeds $N_\text{str}$, the algorithm stops leaving other groups are left empty.

\begin{algorithm}
\small
  \centering
  \caption{\gls{UE} stream selection based on large-scale propagation parameters.}
  \begin{flushleft}
  \textbf{Necessary condition:} $K< N_\text{str}$\\
  \textbf{Inputs:} $\mathcal K$, $\mathcal P$, $N_\text{UE}$, $N_\text{str}$, $\beta_{mk}, \bs R_{k} \ \ \forall m,k$,\\ 
  \ \ \ \ \ \ \ \ \ \ $\theta_{\xi}^{\text{good},1},\hdots, \theta_{\xi}^{\text{good},N_\text{UE}-1},\theta_{\beta}^{\text{good},1},\hdots, \theta_{\beta}^{\text{good},N_\text{UE}-1}$.\\
  $\forall k$ \textbf{Compute strength and \gls{UE} conditioning: } 
  $$\beta_k=\text{Trace}(\bs R_k) \text{ and }\xi_k=\lambda_k^\text{max}/\lambda_k^\text{min} \text{, where }$$ 
  $$\lambda_k^\text{max}=\max\{\text{eigs}(\bs R_k)\},\lambda_k^\text{min}=\min\{\text{eigs}(\bs R_k)\},$$
   $L=K$ (Total umber of streams allocated. At least 1 per \gls{UE}).\\
   $\mathcal K^{(1)}=\hdots=\mathcal K^{(N_\text{UE})}=\emptyset$ \\
   $\mathcal K^{(0)}_\text{rest}=\mathcal K$\\
   $\mathcal P_1=\hdots=\mathcal P_{N_\text{UE}}=\emptyset$ \\
   $\mathcal P^{(0)}_\text{rest}=\mathcal P$\\
   $g=1$\\
   \textbf{while } $(g<N_\text{UE})$ \textbf{ and } $(L<N_\text{str})$ \textbf{ do }\\
     \ \ \ \ \ \ \ \ $\hat{\mathcal K}= \left\{k \in \mathcal K^{(g-1)}_\text{rest}: \beta_k > \theta_{\beta}^{\text{good},g} \text{ and }  \xi_k<\theta_{\xi}^{\text{good},g}  \right\}$\\
    \ \ \ \ \ \ \ \ \textbf{if } $|\hat{\mathcal K}|< (N_\text{str}-L)$ \textbf{ then }\\
    \ \ \ \ \ \ \ \ \ \ \ \ \ \ \ \ $\mathcal K^{(g)}_\text{rest}=\hat{\mathcal K}$\\
    \ \ \ \ \ \ \ \ \ \ \ \ \ \ \ \ $\mathcal K^{(g)}=\mathcal K^{(g-1)}_\text{rest} - \hat{\mathcal K}$\\
   
    \ \ \ \ \ \ \ \ \textbf{else }\\
    \ \ \ \ \ \ \ \ \ \ \ \ \ \ \ \ $\hat{\mathcal K}_\text{tmp}=\{k \in \hat{\mathcal K} \text{ with } (N_\text{str}-L)-|\hat{\mathcal K}| \text{ largest } \beta_k\}$\\
  \ \ \ \ \ \ \ \ \ \ \ \ \ \ \ \ $\mathcal K^{(g)}_\text{rest}=\hat{\mathcal K}_\text{tmp}$\\
    \ \ \ \ \ \ \ \ \ \ \ \ \ \ \ \ $\mathcal K^{(g)}=\mathcal K^{(g-1)}_\text{rest} - \hat{\mathcal K}_\text{tmp}$\\
    \ \ \ \ \ \ \ \ \textbf{end }\\
     \ \ \ \ \ \ \  $L=L+| K^{(g)}_\text{rest}|$\\
    \ \ \ \ \ \ \ \ $|\mathcal P_g|=\text{round}\left\{\frac{|\mathcal K^{(g)}||\mathcal P_{g-1}|}{|\mathcal K^{(g-1)}|} \right\}$\\
    \ \ \ \ \ \ \ \ $\mathcal P_g=\{|\mathcal P_g|  \text{ columns of } \mathcal P_\text{rest}^{(g-1)}\}$\\
    \ \ \ \ \ \ \ \ \textbf{if } $\mathcal P_g=\emptyset$ (Not enough pilot symbols) \\
    \ \ \ \ \ \ \ \ \ \ \ \ \ \ \ \ $\mathcal K^{(g)}=\emptyset$\\
    \ \ \ \ \ \ \ \ \ \ \ \ \ \ \ \ $\mathcal K^{(g)}_\text{rest}=\emptyset$\\
    \ \ \ \ \ \ \ \ \ \ \ \ \ \ \ \ \textbf{break} from while loop\\
    \ \ \ \ \ \ \ \ \textbf{end }\\
    
    \ \ \ \ \ \ \ \ $g=g+1$\\
    \textbf{end }\\
  \textbf{Output: } \gls{UE} groups $\mathcal K^{(1)}, \hdots, \mathcal K^{(N_\text{UE})}$\\
      \ \ \ \ \ \ \ \ \ \ \ \ Pilot groups $\mathcal P_{1}, \hdots, \mathcal P_{N_\text{UE}}$.\\
  \end{flushleft}
  \label{EE_efficient_algorithm}
\end{algorithm}

\subsection{Adaptive \gls{TRP} (de)activation}
\label{sec:adaptive}
The adaptive \gls{TRP} (de)activation proposed in this paper targets the maximization of the overall network \gls{EE} subject to a constraint on the achieved \gls{SE}.
Given the baseband processing introduced in Section \ref{sec:BB},
the only degree of freedom left to tune the network performance is the setting of the connectivity matrix $\boldsymbol C$,
which is initially subject to constraints, 
if scalability needs to be guaranteed.
Formally,
fixing a prescribed number of active \glspl{TRP},
the design of $\boldsymbol C$ is governed by the following optimization problem
\begin{equation}
    \begin{split}
    & \argmax_{\boldsymbol C}{EE} \\
    \quad &\text{ subject to}\\
    \quad & \quad \quad \quad  \|\boldsymbol c_{[m]}\|_1\leq K_\text{TRP} \ \ \ \forall m \in \mathcal M\\
    \quad & \quad \quad \quad  \|\boldsymbol c^{[k]}\|_1\geq 1 \ \ \ \forall k \\
    \quad & \quad \quad \quad f(\{R_k\})\geq R_0 \ \ \ \forall k, \\
    \end{split}
    \label{optim}
\end{equation}
where the first restriction ensures the scalability of the network,
the second ensures that no \gls{UE} is left without connection,
and the third establishes a general rate restriction, 
with $f(\cdot)$ denoting an arbitrary function of the achievable \gls{SE} experienced by all \glspl{UE}.
Popular choices for this function could be the average, the minimum or a certain percentile of the \glspl{UE}' rates.
Note that this optimization problem implicitly involves determining how many \glspl{TRP} are active ($M$), 
and which ones.
Some remarks are in place:
\begin{enumerate}
\item 
The minimum rate constraint unavoidably implies that problem \eqref{optim} may not be feasible, 
in which case,
an outage occurs.
\item Any \gls{TRP} $m$, for which $\|\boldsymbol c_{[m]}\|_1=0$, belongs to $\mathcal M_T \setminus \mathcal M$, 
and thus the \gls{TRP} power consumption is reduced in accordance to \eqref{eq:powermodelAAU}.
\item 
The \gls{CF-M-MIMO} scheme can be turned into a conventional co-located (cellular) \gls{M-MIMO} by activating all \glspl{TRP}, 
i.e. $\mathcal M=\mathcal M_T$, 
and setting the $K$ columns of $\boldsymbol C$ as
\begin{equation}
  \boldsymbol c^{[k]}=[\underbrace{0 \ldots 0}_{m_k-1} \ 1 \ \underbrace{0 \ldots 0}_{M-m_k+1}]^T,
\end{equation}
where $m_k=\argmax_{m}\beta_{mk}$, 
in which case every \gls{TRP} solely serves the \glspl{UE} within its cell (here it is assumed that \gls{UE}-association is conducted on the basis of minimum propagation losses, 
or equivalently, maximum received signal strength).
\item 
We note the absence of a per-\gls{TRP} maximum power constraint in problem \eqref{optim}. 
The reason is that by adhering to the fractional power allocation in \eqref{power_dl}, 
such restriction is guaranteed to be fulfilled.
\end{enumerate}

Problem \eqref{optim} is a constrained binary optimization problem,
whose solution requires an exhaustive search over all candidate $\boldsymbol C$ matrices.
Since this approach quickly becomes computationally impractical,
even for moderate values of $M_T, M$ or $K$,
alternative solutions need to be sought.

We first propose an heuristic to approach the solution of \eqref{optim}
that relies on the \gls{DCF} from \cite{demir2021}, 
while ensuring:
\begin{enumerate}
  \item
  $\mathcal M=\mathcal M_T$.
  \item
  Each \gls{TRP} $m \in \mathcal M$ serves the $K_{TRP}$ \glspl{UE} experiencing the largest large-scale gains.
  That is,
  the connectivity vector of the $m$th \gls{TRP} $\boldsymbol c_\text{[m]}$ is defined as
        \begin{equation}
            \boldsymbol c_{mk}=\left\{\begin{array}{cc}
                             1 & \text{ if } \beta_{mk}\in \mathcal S_m^{K_{TRP}} \\
                             0 & \text{ otherwise }
                           \end{array}\right.,
            \label{maxBetas}
        \end{equation}
  where $S_m^{K_{TRP}}$ is the set formed by the largest $K_{TRP}$ entries of all large-scale gains involving the $m$th \gls{TRP}.%, $\{\beta_{m1} \ldots \beta_{mK}\}$.
  \item
  For any \gls{UE} $k'$,
  for which  $\|\boldsymbol c^{[k']}\|_1=0$ (\emph{orphan} \gls{UE}),
  its strongest \gls{TRP} is identified by $m_{k'}=\argmax_{m}\beta_{mk'}$,
  and the connectivity of \gls{TRP} $m_{k'}$ is modified by setting:
        \begin{equation}
            c_{m_{k'}k'}=1 \text{ and } c_{m_{k'}i}=0,
            \label{orphan}
        \end{equation}
       where $i=\argmin_{i} S_{m_{k'}}^{K_{TRP}}$.
       This step trades the weakest \gls{UE} selected by \gls{TRP} $m_{k'}$ by the orphan \gls{UE}.
\end{enumerate}
This heuristic proceeds in a greedy fashion switching off at each step the \gls{TRP} that maximizes the \gls{EE},
while satisfying the constraints in \eqref{optim}.
This method, however, is still computationally intensive as it involves repeatedly calculating \eqref{CMMSE}.
% This procedure is repeated until the \gls{EE} decreases or any of the constraints in \eqref{optim} is unfulfilled. Unfortunately, this search is nevertheless computationally intensive as it entails the computation of all \gls{UE} rates using \eqref{eq:rate}, which in turn implies the repeated calculation of $K$ matrix inverses of dimensions $MN_{TRP}\times MN_{TRP}$.

Striving for a simpler solution to Problem \eqref{optim},
we introduce here a lower-complexity greedy alternative that determines the \gls{TRP} to be switched off by only relying on the large-scale propagation losses rather than on the achievable rates.
The algorithm proposed builds on previous proposals \cite{femenias2020,riera-palou2021},
but noting that, unlike in classical (non-regular) \gls{CF-M-MIMO} deployments, 
here $K\gg{M}$.
The algorithm proceeds in a greedy fashion aiming at switching off at each step (indexed by $l$) the \gls{TRP} whose large-scale losses has the least performance degradation for the overall \gls{UE} population.

In order to formalize this strategy,
summarized in Alg.~\ref{EE_efficient_algorithm},
let us define set $\boldsymbol \beta_{\mathcal K}=[\bs\beta_{1} \ldots \bs\beta_{K}]$ with $\bs \beta_{k}=[\beta_{1k} \ldots \beta_{Mk}]$ representing the vector containing all the large-scale propagation gains from the $k$th \gls{UE} to all active \glspl{TRP}. At each step $l$ of the algorithm, the goal is to reduce the set of active \glspl{TRP} from $\mathcal M^{(l-1)}$ to $\mathcal M^{(l)}$ by removing the \gls{TRP} $m'$ that has less performance impact on the \glspl{UE}. Further refining the notation, when beginning step $l$ of the algorithm, the \gls{UE}-specific signature vector is defined by $\boldsymbol \beta_k^{(l)}=[\beta_{1k} \ldots \beta_{M^{(l-1)}k}]$, which comprises the gains to the set of active \glspl{TRP} $\mathcal M^{(l-1)}$. In order to identify the \gls{TRP} $m'$ to switch off, \glspl{UE} are clusterized into $M^{(l)}-1$ disjoint sets on the basis of their signature-vector $\boldsymbol \beta_k^{(l)}$. Popular clustering techniques, such as $k$-means,
can be used to partition at each step the set of all \gls{UE} into clusters $\mathcal K^{(l)}$,
i.e., $\mathcal K^{(l)}=\{\mathcal K_1, \ldots, \mathcal K_{M^{(l)}}\}$.
This clustering step will naturally tend to group together those \glspl{UE} that lie geographically close \cite{riera-palou2018}.
Aside from a cluster index,
$k$-means also returns the centroid of each cluster $c$,
i.e., $\bar{\boldsymbol \beta}_{\mathcal K^{(l)}}(c)$,
with $c \in (1,M^{(l)})$.

At the $l$th step, the algorithm proceeds to first select (in an ordered manner) the \glspl{TRP} exhibiting the minimum propagation losses to each of the $M^{(l)}$ centroids. In a second step, after removing the already selected \glspl{TRP} from the set of selectable ones, the procedure is repeated. That is, the algorithm again orderly selects the \glspl{TRP}, out of the remaining ones, whose propagation losses to each
of the $M^{(l)}$ centroids is minimum. This procedure is repeated until
the number of selected \glspl{TRP} is equal to $M^{(l)}=M^{(l-1)}-1$ .
Finally, the \gls{EE} of the network with such configuration is estimated using \eqref{eq:EE}.
This procedure is repeated until the maximum \gls{EE} is identified,
while satisfying the constraints in \eqref{optim}, 
or there are no more \glspl{TRP} to be turned off.
%In the case of lightly loaded networks,
%where $K\leq M$,
%the k-means clustering can be spared,
%and the candidate \gls{TRP} to be switched off is simply the \gls{TRP} experiencing the lowest %large-scale gain among all \glspl{UE}.
%We note that different choices are possible for the distance governing the k-means clustering $d(\boldsymbol x,\boldsymbol y)$ and the centroid-summarizing metric $\gamma_c$. The specific choices are given in the numerical results section.  The whole procedure is summarized in Alg.~\ref{EE_efficient_algorithm}.
\begin{algorithm}
\small
  \centering
  \caption{\emph{Greedy-efficient} adaptive algorithm based on large-scale propagation losses.}
  \begin{flushleft}
  \textbf{Inputs:} $R_0,\mathcal M_T, K_\text{TRP}, \beta_{mk} \ \ \forall m,k \in \mathcal M_T$.\\
  \textbf{Initialization: } Apply \gls{DCF} calculating $\mathcal M=\mathcal M_T$ and setting $\boldsymbol c_\text{[m]}$ for all $m$ using \eqref{maxBetas} and \eqref{orphan}.\\ 
  Estimate \glspl{UE}' rates $R_k \ \ \forall k$, using \eqref{eq:rate}.\\
  Estimate baseline energy efficiency, EE$^{(0)}$, using \eqref{eq:EE}.\\
  $\mathcal M^{(0)}=\mathcal M_T$, $M^{(0)}=M_T$,  $\boldsymbol {\beta}_{\mathcal K}^{(0)}=\boldsymbol \beta_{\mathcal K}$. $l=1$. $m'=\emptyset$\\
  \textbf{While } ($f(\{R_k\})\geq R_0$) and ($l<M$) \\
     \ \ \ \ \ \ \ \textbf{if } $K>M^{(l-1)}$ \textbf{then }\\
     \ \ \ \ \ \ \ \ \ \ \ \ \ \ $M^{(l)}=M^{(l-1)}-1$,\\
     \ \ \ \ \ \ \ \ \ \ \ \ \ \ Use k-means to construct $M^{(l-1)}$ \gls{UE} clusters\\ \ \ \ \ \ \ \ \ \ \ \ \ \ \ $\mathcal K^{(l)}=\{\mathcal K_1, \ldots, \mathcal K_{M^{(l-1)}}\}$ with centroids\\
     \ \ \ \ \ \ \ \ \ \ \ \ \ \ \ \ \ \ \ $\bar{\boldsymbol \beta}_{\mathcal K^{(l)}}=\{\bar{\boldsymbol \beta}_{\mathcal K_1},\ldots,\bar{\boldsymbol \beta}_{\mathcal K_{M^{(l-1)}}}\}$.\\
          \ \ \ \ \ \ \ \ \ \ \ \ \ \ $\mathcal J$=Orderly collect from set $\bar{\boldsymbol \beta}_{\mathcal K^{(l)}}$ the $\mathcal M^{(l)}$ \glspl{TRP}\\
          \ \ \ \ \ \ \ \ \ \ \ \ \ \ \ \ \ \ with largest virtual propagation gains.\\
          \ \ \ \ \ \ \ \ \ \ \ \ \ \ $m'=\mathcal M^{(l-1)}\setminus \mathcal J$\\
     \ \ \ \ \ \ \ \textbf{else }\\
          \ \ \ \ \ \ \ \ \ \ \ \ \ \ $\mathcal J$=Orderly collect from set ${\boldsymbol \beta}_{\mathcal K^{(0)}}$ the $\mathcal M^{(l)}$ \glspl{TRP}\\
          \ \ \ \ \ \ \ \ \ \ \ \ \ \ \ \ \ \ with largest large-scale gains.\\
          \ \ \ \ \ \ \ \ \ \ \ \ \ \ $m'=\mathcal M^{(l-1)}\setminus \mathcal J$\\

     \ \ \ \ \ \ \ \textbf{end if}\\

     \ \ \ \ \ \ \ Update energy efficiency, EE$^{(l)}$.\\
     \ \ \ \ \ \ \ \textbf{if } (EE$^{(l)}<$EE$^{(l-1)}$) \textbf{then }\\
     \ \ \ \ \ \ \ \ \ \ \ \ \ \ Break from while loop\\
     \ \ \ \ \ \ \ \textbf{else }\\
     \ \ \ \ \ \ \ \ \ \ \ \ \ $l=l+1$.\\
     \ \ \ \ \ \ \ \ \ \ \ \ \ \ Update connectivity matrix $\mathbf C$ by removing the $m'$th row.\\
     \ \ \ \ \ \ \ \ \ \ \ \ \ Update active \glspl{TRP}: $\mathcal M^{(l)}=\mathcal M^{(l-1)} \setminus m'$.\\
     \ \ \ \ \ \ \ \textbf{end if }\\
  \textbf{end while }.\\
  \textbf{Output: } Connectivity matrix $\mathbf C$.\\
  \end{flushleft}
  \label{EE_efficient_algorithm}
\end{algorithm}

\section{Numerical results}
\label{sec:results}

As discussed in Section \ref{sec:model},
the scenario considered is inspired by the \gls{eMBB} dense urban scenario test environment in \cite{itu2017},
depicted in Fig.~\ref{topology},
where in this case,
$L=7$, $S=3$ and thus $M_T=21$.
The inter-site distance is set to 200 m.
Wrap-around is applied to the target coverage area to eliminate boundary effects from the simulation results.
Every \gls{TRP} is equipped with a full digital array of $N_{TRP}=64$ antenna elements,
configured as an $16\times 4$ \gls{URA},
with each antenna element connected to a transceiver chain.
%Following the model in \cite[Chapter 2]{bjornson2017}, $\bs R_{mk}^\text{TRP}$ and $\bs R_{mk}^\text{UE}$ are calculated assuming an \gls{AoA} that follows a Gaussian distribution around the nominal angles $\vartheta_{mk}^\text{TRP}$ and $\vartheta_{mk}^\text{UE}$, respectively.
%David: I have moved this to the system model.
The third constraint in Problem \eqref{optim} has been imposed on the average rate as per the IMT-2020 requirements,
i.e., the EE maximization should fulfill $E\{R_1,\hdots,R_K\}\geq R_0$ with $R_0=100$ Mbps.
Table \ref{table_param} presents the rest of parameters used.

\begin{table}[!t]
\vspace{0.25cm}
\renewcommand{\arraystretch}{1.1}
\caption{\small Summary of default simulation parameters}
\label{tab:default_parameters}
\centering
\begin{tabular}{l|c}
\hline
\bfseries Parameter (symbol) & \bfseries Value\\
\hline
\footnotesize {Carrier frequency ($f_c$):} & \footnotesize {2 GHz}\\
\footnotesize {Bandwidth ($B$)} & \footnotesize {100 MHz}\\
\footnotesize {Fixed TRP power ($P_{fix}$)} & \footnotesize {500 W}\\
\footnotesize {Maximum radiated TRP power ($P_{out}$)} & \footnotesize {240    W}\\
\footnotesize {CPU power consumption ($P^\text{CPU,fix},P^\text{CPU,pre}$)} & \footnotesize {5 W, 0.1 W/Gbps}\\
\footnotesize {Fronhtaul power consumption ($P_m^\text{FH,fix},P_m^\text{FH,var}$)} & \footnotesize {0.825 W, 0.01 W}\\
\footnotesize {Fractional power allocation factor ($\upsilon$)} & \footnotesize {-0.5}\\
\footnotesize {TRP $/$ UE antenna height ($h_{TRP} / h_{UE}$)} & \footnotesize {25 m $/$ 1.65 m}\\
%\footnotesize {MS antenna height $h_{MS}$} & \footnotesize {1.65 m}\\
\footnotesize {Coherence interval length ($\tau_c$)} & \footnotesize {200 samples}\\
\footnotesize {Training phase length ($\tau_p$)} & \footnotesize {20 samples}\\
\footnotesize {Pathloss parameters ($\iota_0$, $\alpha$, $\sigma_{\chi}$)} &  \footnotesize{30, 3.67, 4} \\
%\footnotesize {      - Case LOS, $10 m < d_{mk} \leq 150 m$} & \footnotesize{34, 2.2, 3} \\
%\footnotesize {      - Case LOS, $d_{mk} > 150 m$} & \footnotesize{-5.17, 4, 3} \\
%\footnotesize {      - Case NLOS} & \footnotesize{30, 3.67, 4} \\
\footnotesize {Shadow fading decorrelation distance ($d_{\text{dcorr}}$}) & \footnotesize {9 m}\\
\footnotesize {Shadow fading correlation among \glspl{TRP}} & \footnotesize {0.5}\\
\footnotesize {Distribution of the AoA deviation} & \footnotesize {$\zeta\sim \mathcal{N}(0,\sigma_\zeta^2)$}\\
\footnotesize {Azimuth Angular standard deviations ($\sigma_\zeta^{A}$)} & \footnotesize{$15^\circ$}\\
\footnotesize {Elevation Angular standard deviations ($\sigma_\zeta^{E}$)} & \footnotesize{$10^\circ$}\\
\hline
\end{tabular}
\label{table_param}
  \vspace{-0.65cm}
\end{table}

\subsection{Benefits of cell-free processing}

\begin{figure}[!t]
\vspace{0.1cm}
      \begin{center}
       \includegraphics[width=.96\linewidth]{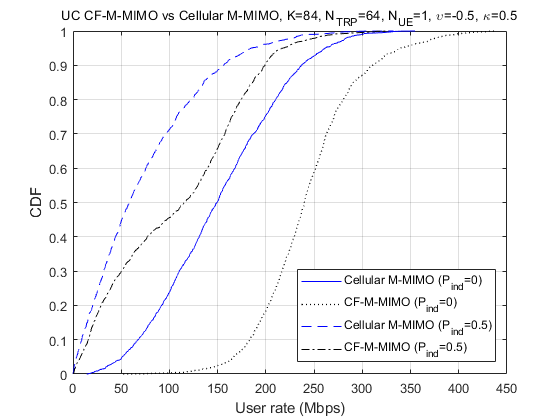}
      \end{center}
      \caption{CF-M-MIMO vs Cellular \gls{UE} rate CDF comparison when $K=84$ single-antenna \glspl{UE} ($N_{UE}=1$).}
      \label{Mbps84}
      \vspace{-0.5cm}
\end{figure}

The first question that must be answered is whether cell-free processing offers a significant throughput advantage over classical cellular-based M-MIMO processing when applied over a regular topology.
Towards this end,
Fig.~\ref{Mbps84} %and \ref{Mbps168}
depicts the \glspl{CDF} of the \gls{UE}'s rates when $K=84$ \glspl{UE}. % and $K=168$, respectively.
Two distinct situations are considered:
One where all \glspl{UE} are outdoors ($P_{ind}=0.0$) and another one where half of the \glspl{UE} are indoors ($P_{ind}=0.5$).
In both cases,
we follow the guidelines in \cite{itu2017} that presume that \glspl{UE} are evenly distributed among the 21 hexagonal cells,
and uniformly distributed within each cell.
For the case of all-outdoor \glspl{UE},
results show how CF-M-MIMO processing offers a considerable advantage.
For the 10\%-tile of worst \glspl{UE},
the CF-M-MIMO reaches 175 Mbps,
whereas the cellular M-MIMO network falls below 70 Mbps,
a 2.5$\times$ increase. %(about 10 dB gain in SINR in Fig.~\ref{SINR84}).
When considering that some of the \glspl{UE} are indoors,
CF-M-MIMO keeps offering a substantial advantage for the best performing \glspl{UE}.
However, this benefit reduces.
When considering the 10\%-tile of worst \glspl{UE},
which invariably corresponds to indoor \glspl{UE} suffering the extra 20 dB loss due to wall propagation,
CF-M-MIMO still offers a 1.6$\times$ gain w.r.t. cellular M-MIMO (from 15 Mbps to 9 Mbps).
The change in the silhouette of the CF-M-MIMO curve when $P_{ind}=0.5$ is due to its different ability to serve indoors and outdoors \glspl{UE}.

\begin{figure}
      \begin{center}
       \includegraphics[width=.96\linewidth]{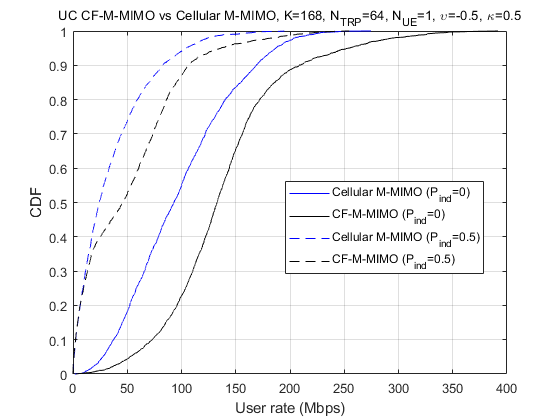}
      \end{center}
      \caption{CF-M-MIMO vs Cellular \gls{UE} rate CDF comparison when $K=168$ single-antenna \glspl{UE} ($N_{UE}=1$).}
      \label{Mbps168}
     \end{figure}

Turning now our attention to the case of a more heavily loaded network, 
shown in Fig.~\ref{Mbps168} (8 \glspl{UE} per hexagon), 
it is clear that the sharing of the \glspl{TRP} available transmit power among more \glspl{UE} together with the larger amount of interference causes the per-\gls{UE} rates to drastically fall. In fact, note that $K=168$ \glspl{UE} is beyond the optimum load point, 
as shown in Fig.~\ref{sumratediffK}. 
Nonetheless, most of the insights drawn when $K=84$ \glspl{UE} still hold with the CF-M-MIMO approach doubling the rate achieved by the worse 10$\%$ of \glspl{UE} in the fully-outdoor scenario. Unfortunately, note how both cellular and CF-M-MIMO architectures have difficulties in handling indoor \glspl{UE}, 
as evidenced by the dashed lines in Fig.~\ref{Mbps168}, 
thus reinforcing the importance of properly dimensioning the network to the traffic needs.

\subsection{Benefits of adaptive UE stream management}

\begin{figure}[!t]
     \begin{center}
      \includegraphics[width=.96\linewidth]{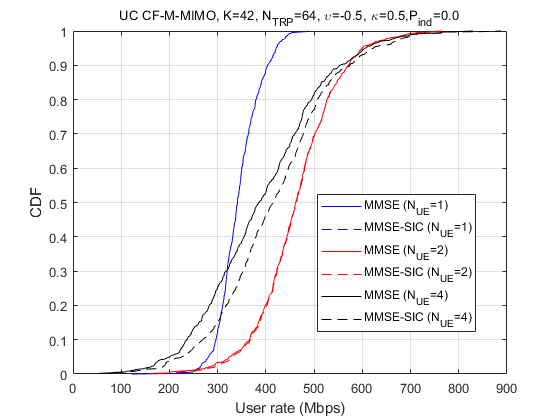}
     \end{center}
     \caption{CDF of \gls{UE} rate for CF-M-MIMO when $K=42$ \glspl{UE} and varying number of the \gls{UE} antennas.}
     \label{varNue}
\end{figure}

As mentioned in the introduction, 
most \gls{CF-M-MIMO} literature focuses on single-antenna \glspl{UE}, 
thus obviating the fact that currently deployed \gls{5G} networks must deal with \glspl{UE} equipped with a larger number of antennas. 
In order to assess this condition, 
Fig.~\ref{varNue} shows the rate performance of a \gls{CF-M-MIMO} network with $K=42$ \glspl{UE} equipped with either 1, 2 or 4 antennas, 
and receiving as many data streams as receive antennas. 
This figure depicts the rate \gls{CDF} attained when \glspl{UE} rely on either of the detection schemes introduced in Section \ref{sec:BB} (i.e., \gls{MMSE} or \gls{MMSE}-\gls{SIC}). 
First point to notice is that increasing the number of received streams per \gls{UE} favours the use of the more advanced \gls{MMSE}-\gls{SIC} detector over \gls{MMSE}. 
In particular, there is no difference between the two detectors for $N_\text{UE}=1$, 
a very marginal improvement for $N_\text{UE}=2$ 
and a clear gain for $N_\text{UE}=4$. 
An even more remarkable observation is the relative performance attained for the different values of $N_\text{UE}$. 
Whereas the configuration for $N_\text{UE}=2$ clearly outperforms single-antenna setups ($N_\text{UE}=1$), 
increasing to $N_\text{UE}=4$ is a rather compromising option: 
For the worst \glspl{UE}, 
actually allowing them to receive 4 streams ($N_\text{UE}=4$) degrades their rate performance below the one that can be achieved when only one stream is received. 
Furthermore, only a very limited number of \glspl{UE} (those with extremely favourable channel conditions) are able to actually gain some rate over that achieved when only activating the reception of two data streams. 
Note that, in this case, $N_\text{str}=KN_\text{UE}=42\times 4=168$, which significantly exceeds the predicted maximum multiplexing gain $K_{\bs \Omega}^\text{opt}=150$, 
hence explaining the degraded performance observed for $N_\text{UE}=4$.
These results illustrate the need to apply the \gls{ASM} scheme introduced in Section \ref{sec:adapt_stream}. 

To this end, 
we now consider the case of an scenario with $K=84$ active \glspl{UE} with multi-antenna capabilities defined by $N_\text{UE}=2$. 
Since each \gls{UE} can process up to two streams, 
the \gls{ASM} mechanism introduced in Section V.A must configure thresholds $\theta_{\xi}^{\text{good,1}}$ and $\theta_{\beta}^{\text{good,1}}$, 
which in the results shown next have been fixed to $\theta_{\beta}^{\text{good,1}}=0.5$ and $\theta_{\xi}^{\text{good,1}}=0.9$. 
These values provide a good balance, 
combining an increase in the multiplexing capabilities of the best \glspl{UE}, 
while preserving the rates of the weakest \glspl{UE}. 
We note that these parameters are not overly sensitive and work well on a relatively wide range on numbers.

\begin{figure}[!t]
     \begin{center}
      \includegraphics[width=.96\linewidth]{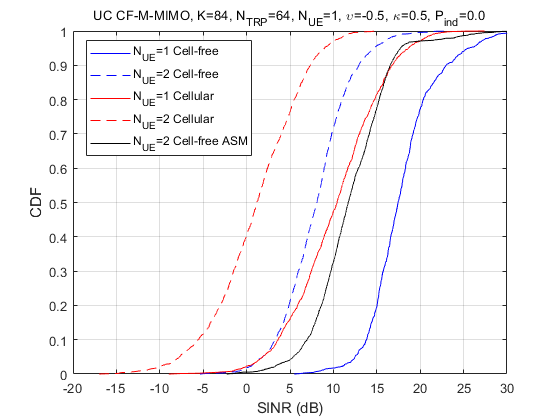}
     \end{center}
     \caption{CDF of per-stream SINR for CF-M-MIMO and Cellular when $K=84$ \glspl{UE}.}
     \label{SINR_adapt}
\end{figure}

Figure \ref{SINR_adapt} compares the \gls{SINR} performance when having one or two antennas at the \glspl{UE} for both, cellular and cell-free topologies. 
All results assume the use of the \gls{MMSE}-\gls{SIC} detector. 
Also note that \gls{SINR} depicted for the two-antenna case actually corresponds to the average \gls{SINR} experienced by the \glspl{UE} on both receive antennas. 
From the figure, 
it can be seen that activating the reception of multiple streams (i.e., using two antennas at the \gls{UE}) always results in an \gls{SINR} degradation mainly due the split of the available transmit power that impinges each data stream. 
Moreover,  Figure \ref{SINR_adapt} also shows how \gls{CF-M-MIMO} attains larger \gls{SINR} values than the corresponding cellular deployments
---a fact that could already be foreseen from the rate results shown in Fig.\ref{Mbps84}. 
Importantly,
these results also indicate the better \gls{SINR} performance when using the \gls{ASM} scheme introduced in Section \ref{sec:adapt_stream}. 

Clearly, the adaptive scheme improves the \gls{SINR} performance over the two-stream counterpart. 
In particular, this improvement is especially clear in the lower \gls{SINR} range. 

\begin{figure}[!t]
     \begin{center}
      \includegraphics[width=.96\linewidth]{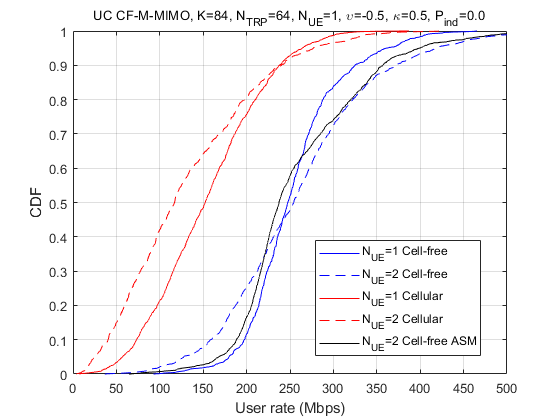}
     \end{center}
     \caption{CDF of \gls{UE} rate for CF-M-MIMO and Cellular when $K=84$ \glspl{UE} for different number of antennas/streams.}
     \label{rate_adapt}
\end{figure}

Despite the hint provided by the \gls{SINR} results, 
it is upon the examination of the rate results in Fig.~\ref{rate_adapt} that most insights can be gathered. 
Noticeably, for both cellular and \gls{CF-M-MIMO}, 
it can be observed how, 
for the lowest rate region, 
the single-stream configuration substantially improves the rate results obtained when using two-stream \glspl{UE}. 
As forecasted in Section \ref{sec:adapt_stream}, 
the increase in pilot sequence re-use (that in turns increases pilot contamination) together the power split cause the two-stream configurations to be severely harmed for the \glspl{UE} that are worse-off. 
In contrast, it can be seen that in the highest rate range, 
characterized by those \glspl{UE} experiencing a strong channel gain and good conditioning, 
the two-stream configuration is superior to the single-stream counterpart. 
%Remarkably, this sort of behaviour is clearly contradictory to the cell-free philosophy that is often hailed as one that avoids cell-edge effects. 
%In fact, it may seems counterintuitive that allowing \glspl{UE} to employ more antennas leads to a clear degradation of the achievable SE experienced by other \glspl{UE}. 
%David: This sentence is innacurate. In reality we are using 2 antenas but 1 data stream, not 1 antenna. Strictly speacking, I think more antennas are good, if they are used optimially. 
It is in this figure where the benefits of the \gls{ASM} technique become most apparent. 
In this specific case, 
the maximum number of data streams has been conservatively set to $N_\text{str}=135(<K_{\bs \Omega}^\text{opt})$. 
Note in Fig.~\ref{rate_adapt} how for the lowest rate range, 
the \gls{ASM} mechanisms basically assigns a single data stream to those \glspl{UE} that are worse off, 
while it assigns two data streams to those \glspl{UE} that are experiencing good propagation conditions. 
In particular, the 10\% of the worst \glspl{UE} only suffer from a degradation of around 3-4\% with respect to that achieved when using a single data stream (from 197 to 188 Mbps),
rather than the 15\% degradation experienced when all \glspl{UE} are allowed to use two data streams (from 197 Mbps to 163 Mbps). 
Complementing these results, note how that for the best \glspl{UE}, 
the \gls{ASM} approach manages to reap off most of the benefits brought by multiple receive antennas at the \gls{UE} side by allowing those \glspl{UE} to receive two data streams.
%David: It may be good to quantify the gain with respect to the case where we force a single data stream per UE

\subsection{Benefits of adaptive \gls{TRP} (de)activation}

Having established the benefits CF-M-MIMO processing can bring to a regular topology,
we address now the maximization of the network gls{EE} through the application of the selective \gls{TRP} activation algorithm introduced in Section \ref{sec:adaptive}.
To fully appreciate the benefits of the greedy \gls{TRP} switch-off,
we assume the presence of two hotspots in the coverage area serving $K=84$ \glspl{UE}.
In particular,
and fixing the central tower in Figure~\ref{topology} to be situated at $(0,0)$ m,
the two hostspots centres are (arbitrarily) located at $(-100, 57.75)$ m and $(200, -57.75)$ m (solid stars in Fig.~\ref{topology}),
and following the guidelines in \cite{3GPP17b},
2/3 of the \glspl{UE} are thrown at either one of the two hotspot areas (within a 40 m radius of the hotspot central location),
while the remaining third is uniformly deployed throughout the coverage area.
In terms of energy consumption,
we consider the most conservative fixed-power reduction factor, $\xi=0.3$, corresponding to symbol shutdown operation \cite{piovesan2022},
presaging that more aggressive sleep modes will result in even larger energy savings.
In order to benchmark the effectiveness of the proposed approach,
and as done in the literature,
the reference used is the random switch-off of \glspl{TRP} \cite{femenias2020}.

\begin{figure}[!t]
     \begin{center}
      \includegraphics[width=.96\linewidth]{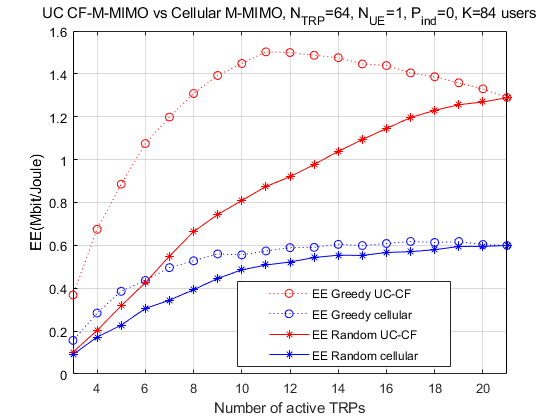}
     \end{center}
     \caption{CF-M-MIMO vs Cellular adaptive switch-off  EE performance when $K=84$ \glspl{UE} with hotspots.}
     \label{EE84hs}
\end{figure}

\begin{figure}[!t]
     \begin{center}
      \includegraphics[width=.96\linewidth]{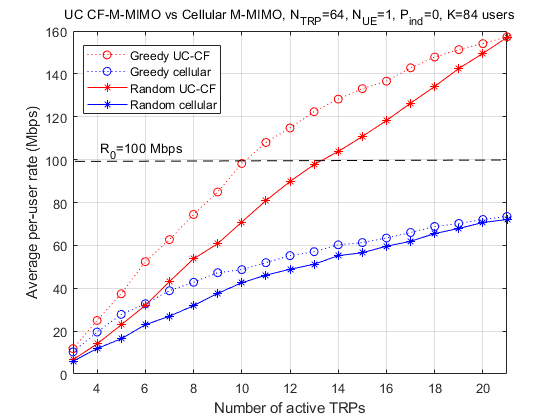}
     \end{center}
     \caption{CF-M-MIMO vs Cellular adaptive switch-off rate performance when $K=84$ with hotspots.}% Circled \gls{UE} depicts a             \gls{CPU}.}% Circled \gls{UE} depicts a \gls{UE} that has been offloaded from the terrestrial to the satellite segment.}
     \label{SE84hs}
\end{figure}

Figures~\ref{EE84hs} and~\ref{SE84hs} respectively depict the average \gls{EE} and \gls{SE} achieved by the cellular M-MIMO and CF-M-MIMO setups when executing either random or greedy switch-off.
%David: The second figure shows rate in Mbps not SE in Mbps/Hz, we need to change the text and figures.
%Antonio: I think it is rate and not SE
The most apparent observation is the clear superiority that the CF-M-MIMO exhibits over cellular in terms of both \gls{EE} and \gls{SE}.
Figure~\ref{EE84hs} depicts the \gls{EE} performance,
and reveals two important facts:
1) the CF-M-MIMO processing significantly outperforms the cellular M-MIMO counterpart; 
2) the greedy switch-off technique provides a clear advantage over the random operation,
with this gain being most visible in the CF-M-MIMO case.
This is because carefully selecting the TRP to be switched-off allows some of the inherent rate loss associated to the switch-off to be compensated through the cooperation of nearby \glspl{TRP}.
In more details, when considering the CF-M-MIMO case,
Alg. 2 achieves an \gls{EE} of 1.5 Mbps/J with just $M=11$ active \glspl{TRP},
whereas the random technique requires the full network ($M=21$) to only reach 1.25 Mbps/J.

Looking now at the rate performance when considering \gls{TRP} shutdown in Figure~\ref{SE84hs},
it is evident that CF-M-MIMO offers a clear advantage over cellular processing.
From this figure,
it can also be seen that Algorithm 2's implementation of the greedy switch-off technique leads to a significant improvement in performance compared to using the random switch-off method.
Cellular M-MIMO is not able to meet the \gls{UE} rate constraint $R_0=100$ Mbps,
also depicted in Fig. \ref{SE84hs},
while \gls{CF-M-MIMO} can comfortably fulfill it by letting on $M\geq10$ \glspl{TRP} when using Algorithm 2 and $M\geq14$ \glspl{TRP} when using random switch-off.
Also note that the switch-off of any \gls{TRP},
independently of the technique in use,
always results in a rate loss.
However, when using Algorithm 2,
this loss takes place more gracefully thanks to the cooperation between \glspl{TRP} and the intelligent \gls{TRP} switch-off. 
Remarkably, operating the network with $M=11$ \glspl{TRP} results in a power consumption reduction of roughly 40\% (from 10.1 kW to 6.1 kW) when compared to a network operating with all \glspl{TRP} active.

\begin{figure}
     \begin{center}
      \includegraphics[width=.96\linewidth]{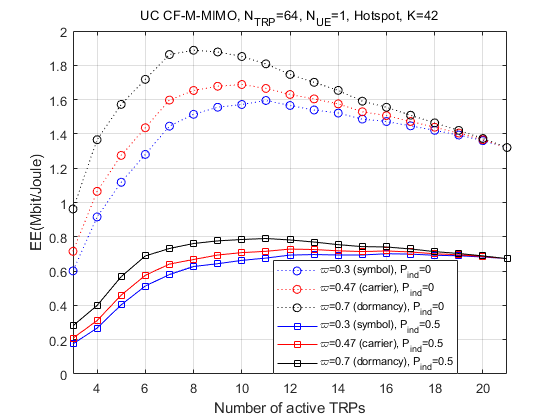}
     \end{center}
     \caption{CF-M-MIMO EE for different fixed-power consumption factors when $K=42$.}
     % Circled \gls{UE} depicts a \gls{CPU}.}% Circled \gls{UE} depicts a \gls{UE} that has been offloaded from the terrestrial to the satellite segment.}
     \label{EEvarXi}
\end{figure}

Figure~\ref{EEvarXi} assesses the performance of the proposed adaptive switch-off under different levels of power consumption while the \glspl{TRP} is in shutdown for the specific case of $K=42$ \glspl{UE} in the network and for different \gls{UE} deployment conditions (i.e., $P_{ind}=0$ or $0.5$). 
In particular, and in accordance to \cite{piovesan2022}, the power reduction coefficient is set to $\varpi=0.3, 0.47$ and $0.7$, corresponding to symbol shutdown, carrier shutdown and dormancy, respectively. 
When focusing on the results for $P_{ind}=0$ ---all \glspl{UE} are outdoor---, 
increasing the power savings during shutdown increases the \gls{EE} and reduces the optimum number of active \glspl{TRP}. 
Note that to maximize \gls{EE}, 
the optimum $M$ drops from 11 to 8 when transiting from symbol shutdown to dormancy savings. 
Although qualitatively the same can be said when $P_{ind}=0.5$, 
clearly the fact of having many indoor \glspl{UE} causes both a drop in \gls{EE} with respect to the all-outdoor \gls{UE} case and the need for a larger number of active \glspl{TRP} to maximize it.

\subsection{Impact of inter-site distance}

\begin{figure}
     \begin{center}
      \includegraphics[width=.96\linewidth]{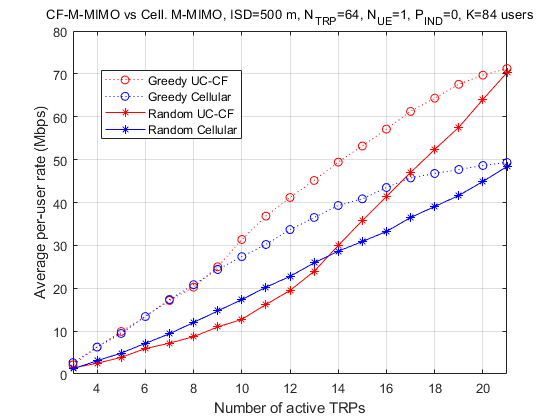}
     \end{center}
     \caption{CF-M-MIMO vs Cellular adaptive switch-off rate performance when $K=84$ with hotspots and \gls{ISD}=500 m.}% Circled \gls{UE} depicts a             \gls{CPU}.}% Circled \gls{UE} depicts a \gls{UE} that has been offloaded from the terrestrial to the satellite segment.}
     \label{ISD500}
\end{figure}

Finally, and in order to illustrate some of the of subtleties to be aware when implementing cell-free schemes, 
Fig.~\ref{ISD500} re-evaluates the results in Fig.~\ref{SE84hs}, 
but now considering an \gls{ISD} of 500 meters. 
Note that this change implies more than quadrupling the coverage area while keeping the same number of \glspl{TRP} and transmit power. 
The first important point to note is the clear degradation in terms of per-\gls{UE} rate mainly caused by the larger propagation losses,
which in turn renders the network unable to fulfill the $R_0=100$ Mbps target. 
It is remarkable that the difference between the greedy and random approaches become relatively much larger as the \gls{ISD} becomes larger. 
This is because the switch-off of a \glspl{TRP} makes its \glspl{UE} to be served by other \gls{TRP} that is farther away. 
Consequently, as the greedy technique is more effective in selecting the \gls{TRP} to deactive than the random one,
it causes less harm to the network performance. 
One more peculiar effect to be observed is that, 
for the random case, 
the purely cellular topology outperforms the \gls{CF-M-MIMO} architecture when the number of active \glspl{TRP} falls below 14 \glspl{TRP}. 
The explanation for this effect has to be sought on the condition imposed on the \gls{CF-M-MIMO} scheme that each \gls{TRP} has to serve $K_\text{TRP}$ \glspl{UE}.
When only a few \glspl{TRP} are active, 
 certain \glspl{UE} are served by \glspl{TRP} that are extremely far away and barely provide any gain yet still induce strict power restrictions as shown in \eqref{power_dl}.

\section{Conclusion}

This paper has quantified for the first time the energy and rate benefits of applying \gls{CF-M-MIMO} processing to a state-of-the-art 5G macrocellular \gls{M-MIMO} network,
with large inter-site distances.
It has been shown that the cell-free principle,
even when applied to a regular topology,
brings along large benefits in terms of increased \gls{SE} when compared to conventional multicell \gls{M-MIMO} architectures.
This improvement is most substantial when serving outdoor \glspl{UE}, 
whereas it is more modest when targeting indoor ones.
Building on a recent and realistic 5G power consumption model,
a novel mechanism has been proposed to selectively (and partially) switch on/off \glspl{TRP} that provides major \gls{EE} gains,
often more than doubling that attained under classical cellular processing. 
It has been shown how multi-antenna \glspl{UE} bring in new challenges due to the exacerbation of the pilot contamination phenomenon. 
In response, a novel strategy has been proposed that only allows those \glspl{UE} with good propagation conditions to actually enjoy multi-stream transmission, 
while guaranteeing at the same time  that \glspl{UE} with poor channel conditions do not have their performance degraded due to an indiscriminate exploitation of the \gls{UE} multi-antenna capabilities.
Results show that the inherent performance loss associated to a \gls{TRP} shutdown can be compensated by allowing nearby \glspl{TRP} to cooperate using \gls{CF-M-MIMO} processing.
Future work will concentrate on the assessment of mixed topologies that combine arbitrarily placed \glspl{TRP} equipped with few antennas with a single \gls{M-MIMO}-equipped \gls{BS} in an attempt to capitalize on the benefits of both architectures.

\bibliographystyle{IEEEtran}
\small
\bibliography{huawei}
\end{document}

%% file: acronyms.tex
\newacronym{3GPP}{3GPP}{third generation partnership project}
\newacronym{5G}{5G}{fifth generation}
\newacronym{6G}{6G}{sixth generation}
\newacronym{AP}{AP}{access point}
\newacronym{AoA}{AoA}{angle-of-arrival}
\newacronym{ASO}{ASO}{adaptive switch on-off}
\newacronym{AAU}{AAU}{active antenna unit}
\newacronym{AWGN}{AWGN}{additive white Gaussian noise}
\newacronym{CF-M-MIMO}{CF-M-MIMO}{cell-free massive MIMO}
\newacronym{C2F-M-MIMO}{C$^2$F-M-MIMO}{clustered cell-free massive MIMO}
\newacronym{CB}{CB}{conjugate beamforming}
\newacronym{CPU}{CPU}{central processing unit}
\newacronym{CSI}{CSI}{channel state information}
\newacronym{GNCB}{GNCB}{globally normalized conjugate beamforming}
\newacronym{MF}{MF}{matched filter}
\newacronym{M-MIMO}{M-MIMO}{massive multiple-input multiple-output}
\newacronym{MIMO}{MIMO}{multiple-input multiple-output}
\newacronym{MMSE}{MMSE}{minimum mean square error}
\newacronym{SIC}{SIC}{successive interference cancellation}
\newacronym{MS}{MS}{mobile station}
\newacronym{NCB}{NCB}{normalized conjugate beamforming}
\newacronym{PM}{PM}{pilot matching}
\newacronym{SNR}{SNR}{signal-to-noise ratio}
\newacronym{SPA}{SPA}{slow power allocation}
\newacronym{TDD}{TDD}{time division duplex}

\newacronym{2D}{2D}{two-dimensional}
\newacronym{3D}{3D}{three-dimensional}
\newacronym{ASM}{ASM}{adaptive stream management}
\newacronym{BS}{BS}{base station}
\newacronym{CAP}{CAP}{compress-after-precoding}
\newacronym{CBP}{CBP}{compress-before-precoding}
\newacronym{CCDF}{CCDF}{complementary cumulative distribution function}
\newacronym{CDF}{CDF}{cumulative distribution function}
\newacronym{CoMP}{CoMP}{coordinated multipoint}
\newacronym{CP-MMSE}{CP-MMSE}{centralized partial-MMSE}

\newacronym{C-RAN}{C-RAN}{cloud radio access network}

\newacronym{DC}{DC}{difference of convex}
\newacronym{DCF}{DCF}{dynamic cluster formation}
\newacronym{DS}{DS}{difference of convex}
\newacronym{DoF}{DoF}{degrees of freedom}
\newacronym{eMBB}{eMBB}{enhanced mobile broadband}
\newacronym{EE}{EE}{energy efficiency}
\newacronym{FD}{FD}{full-dimensional}
\newacronym{FDD}{FDD}{frequency division duplexing}
\newacronym{FD-MIMO}{FD-MIMO}{full-dimensional MIMO}
\newacronym{GOPA}{GOPA}{globally optimal power allocation}
\newacronym{ICIC}{ICIC}{inter-cell interference coordination}
\newacronym{iid}{iid}{independent and identically distributed}
\newacronym{ISD}{ISD}{inter-site distance}
\newacronym{LCAP}{LCAP}{layered \gls{CAP}}
\newacronym{LCBP}{LCBP}{layered \gls{CBP}}
\newacronym{LEO}{LEO}{low Earth orbit}
\newacronym{LOS}{LOS}{line-of-sight}
\newacronym{NLOS}{NLOS}{non-line-of-sight}
\newacronym{NOPA}{NOPA}{normalized optimal power allocation}
\newacronym{NR}{NR}{new radio}
\newacronym{OFDM}{OFDM}{orthogonal frequency division multiplexing}
\newacronym{QoS}{QoS}{quality of service}
\newacronym{RRU}{RRU}{remote radio unit}
\newacronym{RHS}{RHS}{right hand side}
\newacronym{RRH}{RRH}{remote radio head}
\newacronym{RRT}{RRT}{radio resource token}
\newacronym{RV}{RV}{random variable}

\newacronym{rms}{rms}{root mean square}
\newacronym{SINR}{SINR}{signal-to-interference-plus-noise ratio}
\newacronym{SE}{SE}{spectral efficiency}
\newacronym{TRP}{TRP}{transmit/reception point}
\newacronym{TRxP}{TRxP}{transmit point}
\newacronym{UE}{UE}{user equipment}
\newacronym{UL}{UL}{uplink}
\newacronym{URA}{URA}{uniform rectangular array}
\newacronym{ZF}{ZF}{zero-forcing}

%% file: Journal_Regular_CF_Massive_SwitchOnOff.bbl
\begin{thebibliography}{10}
\providecommand{\url}[1]{#1}
\csname url@samestyle\endcsname
\providecommand{\newblock}{\relax}
\providecommand{\bibinfo}[2]{#2}
\providecommand{\BIBentrySTDinterwordspacing}{\spaceskip=0pt\relax}
\providecommand{\BIBentryALTinterwordstretchfactor}{4}
\providecommand{\BIBentryALTinterwordspacing}{\spaceskip=\fontdimen2\font plus
\BIBentryALTinterwordstretchfactor\fontdimen3\font minus
  \fontdimen4\font\relax}
\providecommand{\BIBforeignlanguage}[2]{{%
\expandafter\ifx\csname l@#1\endcsname\relax
\typeout{** WARNING: IEEEtran.bst: No hyphenation pattern has been}%
\typeout{** loaded for the language `#1'. Using the pattern for}%
\typeout{** the default language instead.}%
\else
\language=\csname l@#1\endcsname
\fi
#2}}
\providecommand{\BIBdecl}{\relax}
\BIBdecl

\bibitem{jiang2021}
W.~Jiang, B.~Han, M.~A. Habibi, and H.~D. Schotten, ``The road towards {6G}: A
  comprehensive survey,'' \emph{IEEE Open Journal of the Communications
  Society}, vol.~2, pp. 334--366, 2021.

\bibitem{matthaiou2021}
M.~Matthaiou and et~al., ``The road to {6G}: Ten physical layer challenges for
  communications engineers,'' \emph{IEEE Communications Magazine}, vol.~59,
  no.~1, pp. 64--69, 2021.

\bibitem{tataria2021}
H.~Tataria, M.~Shafi, A.~F. Molisch, M.~Dohler, H.~Sj{\"o}land, and
  F.~Tufvesson, ``{6G} wireless systems: Vision, requirements, challenges,
  insights, and opportunities,'' \emph{Proceedings of the IEEE}, vol. 109,
  no.~7, pp. 1166--1199, 2021.

\bibitem{uusitalo2021}
M.~A. Uusitalo, P.~Rugeland, M.~R. Boldi, E.~C. Strinati, P.~Demestichas,
  M.~Ericson, G.~P. Fettweis, M.~C. Filippou, A.~Gati, M.-H. Hamon
  \emph{et~al.}, ``{6G} vision, value, use cases and technologies from european
  {6G} flagship project {H}exa-{X},'' \emph{IEEE Access}, vol.~9, pp.
  160\,004--160\,020, 2021.

\bibitem{marzetta2016}
T.~L. Marzetta, E.~G. Larsson, H.~Yang, and H.~Q. Ngo, \emph{Fundamentals of
  massive {MIMO}}.\hskip 1em plus 0.5em minus 0.4em\relax Cambridge Univ.
  Press, 2016.

\bibitem{lopez2022}
D.~L{\'o}pez-P{\'e}rez, A.~De~Domenico, N.~Piovesan, G.~Xinli, H.~Bao,
  S.~Qitao, and M.~Debbah, ``A survey on {5G} radio access network energy
  efficiency: Massive {MIMO}, lean carrier design, sleep modes, and machine
  learning,'' \emph{IEEE Communications Surveys \& Tutorials}, 2022.

\bibitem{ngo2017}
H.~Q. Ngo, A.~Ashikhmin, H.~Yang, E.~G. Larsson, and T.~L. Marzetta,
  ``Cell-free massive {MIMO} versus small cells,'' \emph{IEEE Transactions on
  Wireless Communications}, vol.~16, no.~3, pp. 1834--1850, 2017.

\bibitem{demir2021}
{\"O}.~T. Demir, E.~Bj{\"o}rnson, and L.~Sanguinetti, ``Foundations of
  user-centric cell-free massive {MIMO},'' \emph{Foundations and
  Trends{\textregistered} in Signal Processing}, vol.~14, no. 3-4, 2021.

\bibitem{mai2022}
T.~C. Mai, H.~Q. Ngo, and L.-N. Tran, ``Energy efficiency maximization in
  large-scale cell-free massive mimo: A projected gradient approach,''
  \emph{IEEE Transactions on Wireless Communications}, vol.~21, no.~8, pp.
  6357--6371, 2022.

\bibitem{femenias2020}
G.~Femenias, N.~Lassoued, and F.~Riera-Palou, ``Access point switch {ON/OFF}
  strategies for green cell-free massive {MIMO} networking,'' \emph{IEEE
  Access}, vol.~8, pp. 21\,788--21\,803, 2020.

\bibitem{garcia2020}
J.~Garc{\'\i}a-Morales, G.~Femenias, and F.~Riera-Palou, ``Energy-efficient
  access-point sleep-mode techniques for cell-free mmwave massive {MIMO}
  networks with non-uniform spatial traffic density,'' \emph{IEEE Access},
  vol.~8, pp. 137\,587--137\,605, 2020.

\bibitem{kanno2022}
I.~Kanno, K.~Yamazaki, Y.~Kishi, and S.~Konishi, ``A survey on research
  activities for deploying cell free massive {MIMO} towards beyond {5G},''
  \emph{IEICE Transactions on Communications}, p. 2021MEI0001, 2022.

\bibitem{kim2022}
T.~Kim, H.~Kim, S.~Choi, and D.~Hong, ``How will cell-free systems be
  deployed?'' \emph{IEEE Communications Magazine}, vol.~60, no.~4, pp. 46--51,
  2022.

\bibitem{mai2020}
T.~C. Mai, H.~Q. Ngo, and T.~Q. Duong, ``Downlink spectral efficiency of
  cell-free massive mimo systems with multi-antenna users,'' \emph{IEEE
  Transactions on Communications}, vol.~68, no.~8, pp. 4803--4815, 2020.

\bibitem{itu2017}
M.~Series, ``Guidelines for evaluation of radio interface technologies for
  {IMT}-2020,'' \emph{Report ITU}, pp. 2412--0, 2017.

\bibitem{checko2014}
A.~Checko, H.~L. Christiansen, Y.~Yan, L.~Scolari, G.~Kardaras, M.~S. Berger,
  and L.~Dittmann, ``Cloud {RAN} for mobile networks-a technology overview,''
  \emph{IEEE Communications surveys \& Tutorials}, vol.~17, no.~1, pp.
  405--426, 2014.

\bibitem{bjornson2020}
E.~Bj{\"o}rnson and L.~Sanguinetti, ``Scalable cell-free massive {MIMO}
  systems,'' \emph{IEEE Transactions on Communications}, vol.~68, no.~7, pp.
  4247--4261, 2020.

\bibitem{3GPP17}
3GPP, ``Study on {3D} channel model for {LTE} {(Release 12)},'' 3GPP TR 36.873
  (version 12.7.0), December 2017.

\bibitem{bjornson2017}
E.~Bj\"{o}rnson, J.~Hoydis, and L.~Sanguinetti, ``Massive {MIMO} networks:
  {Spectral}, energy, and hardware efficiency,'' \emph{Foundations and
  Trends{\textregistered} in Signal Processing}, vol.~11, no. 3-4, 2017.

\bibitem{femenias2019}
G.~{Femenias} and F.~{Riera-Palou}, ``Cell-free millimeter-wave massive {MIMO}
  systems with limited fronthaul capacity,'' \emph{IEEE Access}, vol.~7, pp.
  44\,596--44\,612, 2019.

\bibitem{kay1993}
S.~M. Kay, \emph{Fundamentals of statistical signal processing: estimation
  theory}.\hskip 1em plus 0.5em minus 0.4em\relax Prentice-Hall, Inc., 1993.

\bibitem{bjornson2019}
E.~Bj{\"o}rnson and L.~Sanguinetti, ``Making cell-free massive {MIMO}
  competitive with {MMSE} processing and centralized implementation,''
  \emph{IEEE Transactions on Wireless Communications}, vol.~19, no.~1, pp.
  77--90, 2019.

\bibitem{piovesan2022}
N.~Piovesan, D.~Lopez-Perez, A.~De~Domenico, X.~Geng, H.~Bao, and M.~Debbah,
  ``Machine learning and analytical power consumption models for {5G} base
  stations,'' \emph{IEEE Communications Magazine}, vol.~60, no.~10, pp. 56--62,
  2022.

\bibitem{chen2022}
S.~Chen, J.~Zhang, E.~Bj{\"o}rnson, {\"O}.~T. Demir, and B.~Ai,
  ``Energy-efficient cell-free massive {MIMO} through sparse large-scale fading
  processing,'' \emph{arXiv preprint arXiv:2208.13552}, 2022.

\bibitem{li2016}
X.~Li, E.~Bj{\"o}rnson, S.~Zhou, and J.~Wang, ``Massive mimo with multi-antenna
  users: When are additional user antennas beneficial?'' in \emph{2016 23rd
  International Conference on Telecommunications (ICT)}.\hskip 1em plus 0.5em
  minus 0.4em\relax IEEE, 2016, pp. 1--6.

\bibitem{riera-palou2021}
F.~Riera-Palou, G.~Femenias, J.~Garc{\'\i}a-Morales, and H.~Q. Ngo, ``Selective
  infrastructure activation in cell-free massive {MIMO}: a two time-scale
  approach,'' in \emph{2021 IEEE Globecom Workshops (GC Wkshps)}.\hskip 1em
  plus 0.5em minus 0.4em\relax IEEE, 2021, pp. 1--7.

\bibitem{riera-palou2018}
F.~Riera-Palou, G.~Femenias, A.~G. Armada, and A.~P{\'e}rez-Neira, ``Clustered
  cell-free massive {MIMO},'' in \emph{2018 IEEE Globecom Workshops (GC
  Wkshps)}.\hskip 1em plus 0.5em minus 0.4em\relax IEEE, 2018, pp. 1--6.

\bibitem{3GPP17b}
3GPP, ``Technical specification group radio access network; evolved universal
  terrestrial radio access ({E-UTRA}); further advancements for {E-UTRA}
  physical layer aspects (release 9),'' 3GPP TR 36.814 (version 9.2.0),
  December 2017.

\end{thebibliography}
